\documentclass[12pt,draftcls,peerreviewca,onecolumn]{IEEEtran}
\usepackage{amsmath,epsfig,graphicx,amsbsy,amssymb,citesort,url}
\usepackage{algorithm,algorithmic,fullpage}
\def\real    { \mathbb{R} }
\def \A {\mathcal{M}}

\def \pinv {^\dag} 

\def \x {x}

\def \y {y}

\def \z {z}

\def \alphahat {\widehat{\alpha}}
\def \hhhat {\widehat{H}}
\def \xhat {\widehat{\x}}
\def \hhat {\widehat{h}}
\def \zhat {\widehat{\z}}

\newtheorem{THEO}{Theorem}

\newtheorem{DEFI}{Definition}

\newtheorem{COROLLARY}{Corollary}

\newcommand{\bigo}[1]{\mathcal{O}\left(#1\right)}

\begin{document}

\title{\LARGE\bf Sampling and Recovery of Pulse Streams}
\author{{\em\Large Chinmay Hegde and Richard G. Baraniuk} $\:$\thanks{Email:
\{chinmay, richb\}@rice.edu; Web: dsp.rice.edu/cs. Refer~\cite{cssp,cssp_icassp} for preliminary conference versions of this manuscript. This work was supported by grants NSF CCF-0431150 and CCF-0728867, DARPA/ONR
N66001-08-1-2065, ONR N00014-07-1-0936 and N00014-08-1-1112, AFOSR
FA9550-07-1-0301, ARO MURIs W911NF-07-1-0185 and W911NF-09-1-0383, and the Texas Instruments Leadership University Program.}
\\
Department of Electrical and Computer Engineering \\
Rice University \vspace{-0.0in}
\date{\today}
}

\maketitle

\begin{abstract}

Compressive Sensing (CS) is a new technique for the efficient acquisition of signals, images, and other data that have a sparse representation in some basis, frame, or dictionary.
By sparse we mean that the $N$-dimensional basis representation has just $K\ll N$ significant coefficients; in this case, the CS theory maintains that just $M = \bigo{K \log N}$ random linear signal measurements will both preserve all of the signal information and enable robust signal reconstruction in polynomial time.
In this paper, we extend the CS theory to {\em pulse stream} data, which correspond to $S$-sparse signals/images that are convolved with an unknown $F$-sparse pulse shape.
Ignoring their convolutional structure, a pulse stream signal is $K=SF$ sparse.
Such signals figure prominently in a number of applications, from neuroscience to astronomy.
Our specific contributions are threefold.
First, we propose a pulse stream signal model and show that it is equivalent to an infinite union of subspaces.
Second, we derive a lower bound on the number of measurements $M$ required to preserve the essential information present in pulse streams.
The bound is linear in the total number of degrees of freedom $S + F$, which is significantly smaller than the naive bound based on the total signal sparsity $K=SF$.
Third, we develop an efficient signal recovery algorithm that infers both the shape of the impulse response as well as the locations and amplitudes of the pulses.
The algorithm alternatively estimates the pulse locations and the pulse shape in a manner reminiscent of classical deconvolution algorithms.
Numerical experiments on synthetic and real data demonstrate the advantages of our approach over standard CS.
\end{abstract}


\section{Introduction}
\label{sec:intro}

Digital signal processing systems face two parallel challenges. On the one hand, with ubiquitous  computing power, memory and communication bandwidth, the pressure is on {\em acquisition} devices, such as analog-to-digital converters and digital cameras, to capture signals at ever increasing sampling rates. To date, signal acquisition has  been governed by the Shannon/Nyquist sampling theorem, which states that all the information contained in a signal is preserved if it is uniformly sampled at a rate twice as fast as the bandwidth of its Fourier transform.  On the other hand, to counter the resulting deluge of Nyquist-rate samples, DSP systems must utilize efficient {\em compression} schemes that preserve the essential information contained in the signals of interest. 
Transform compression of a discrete-time signal $x \in \real^N$ involves representing the signal in a suitable basis expansion $x = \Psi \alpha$, with $\Psi$ an $N \times N$ basis matrix, and storing only the $K$ largest basis coefficients. The number of large coefficients in $\alpha$ is known as the {\em sparsity} $K$ of the signal in the basis $\Psi$. For many classes of interesting signals, $K \ll N$, and hence efficient signal compression can be achieved.

An intriguing question can thus be asked: can a system simultaneously attain the twin goals of signal acquisition and compression? Surprisingly, the answer in many cases is {\em yes}. This question forms the core of the burgeoning field of Compressive Sensing (CS)~\cite{Donoho04A,Candes04C}. A prototypical CS system works as follows: a signal $x$ of length $N$ is sampled by measuring its inner products with $M \ll N$ vectors; the output of the sampling system is thus given by the vector $y = \Phi x = \Phi \Psi \alpha$, where $\Phi \in \real^{M \times N}$ is a non-invertible matrix. The CS theory states that with high probability, $x$ can be exactly reconstructed from $y$ provided that ($i$) the elements of $\Phi$ are chosen randomly from subgaussian probability distributions, and ($ii$) the number of samples $M$ is $\bigo{K \log(N/K)}$.
Further, this recovery can be carried out in polynomial time using efficient greedy approaches or optimization based methods~\cite{OMP,BPDN}.

For some applications, there exist more restrictive signal models than simple sparsity that encode various types of inter-dependencies among the locations of the nonzero signal components. 
Recent work has led to the development of CS theory and algorithms that are based on {\em structured sparsity} models that are equivalent to a finite union of subspaces~\cite{modelCS,EldarUSS}. By exploiting the dependencies present among the nonzero coefficients, $M$ can be significantly reduced;
for certain structured sparsity models, with high probability the number of measurements $M$ required for exact recovery is merely $\bigo{K}$ (without the additional logarithmic dependence on the signal length $N$).

Despite the utility of sparsity models, in many real-world sensing applications the assumption of sparsity itself is an oversimplification. For example, a electrophysiological recording of a neuron is often approximated as a series of spikes but can be better modeled as a series of more elongated pulses, the pulse shape being characteristic to the particular neuron. As another example, a high-resolution image of the night sky consists of a field of points (corresponding to the locations of the stars) convolved with the point spread function of the imaging device. Such signals can be modeled as an $S$-sparse {\em spike stream} that have been convolved with an unknown $F$-sparse {\em impulse response} so that the resulting overall sparsity $K = SF$. We call such a signal a {\em pulse streams}. For the compressive sensing and recovery of a pulse stream, the number of measurements $M$ would incur a corresponding multiplicative increase by a factor of $F$ when compared to sensing merely the underlying spike streams; this can be prohibitive in some situations. Thus, it is essential to develop a CS framework that can handle not just sparse signals but also more general pulse streams.

\begin{figure}[!t]
\centering
\begin{tabular}{cc}
{\includegraphics[width=0.3\hsize]{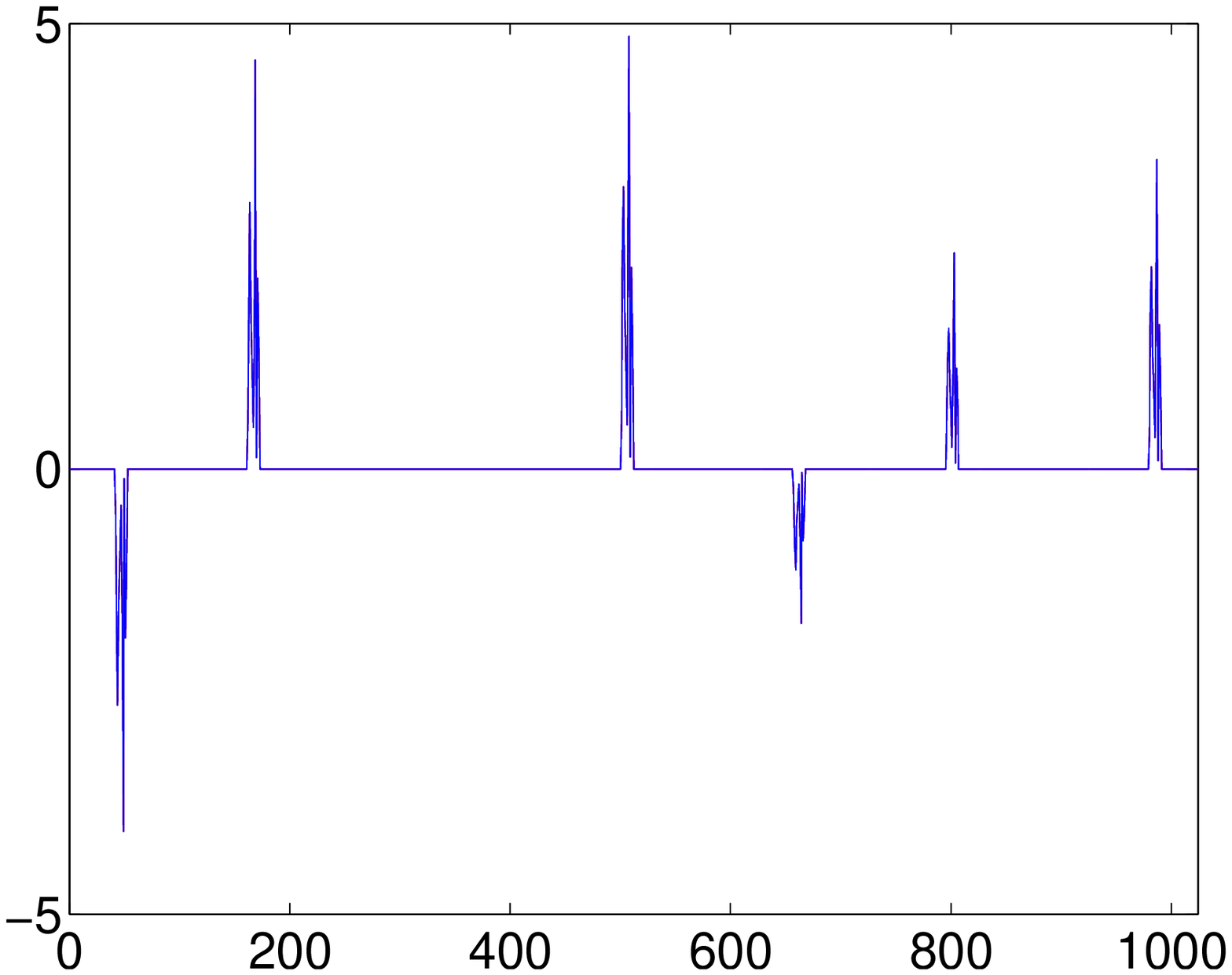}} &
{\includegraphics[width=0.3\hsize]{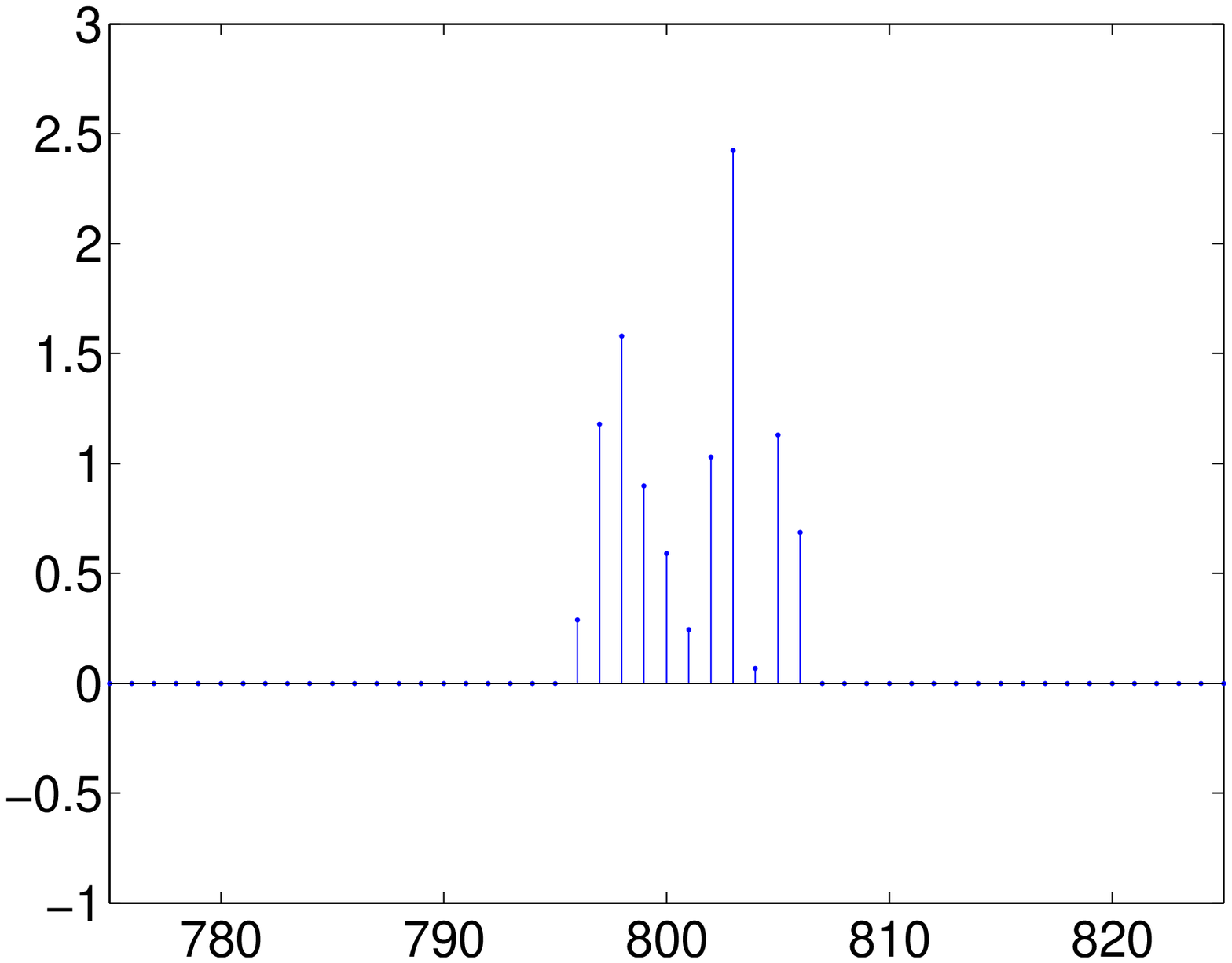}} \\
(a) & (b) \\
{\includegraphics[width=0.3\hsize]{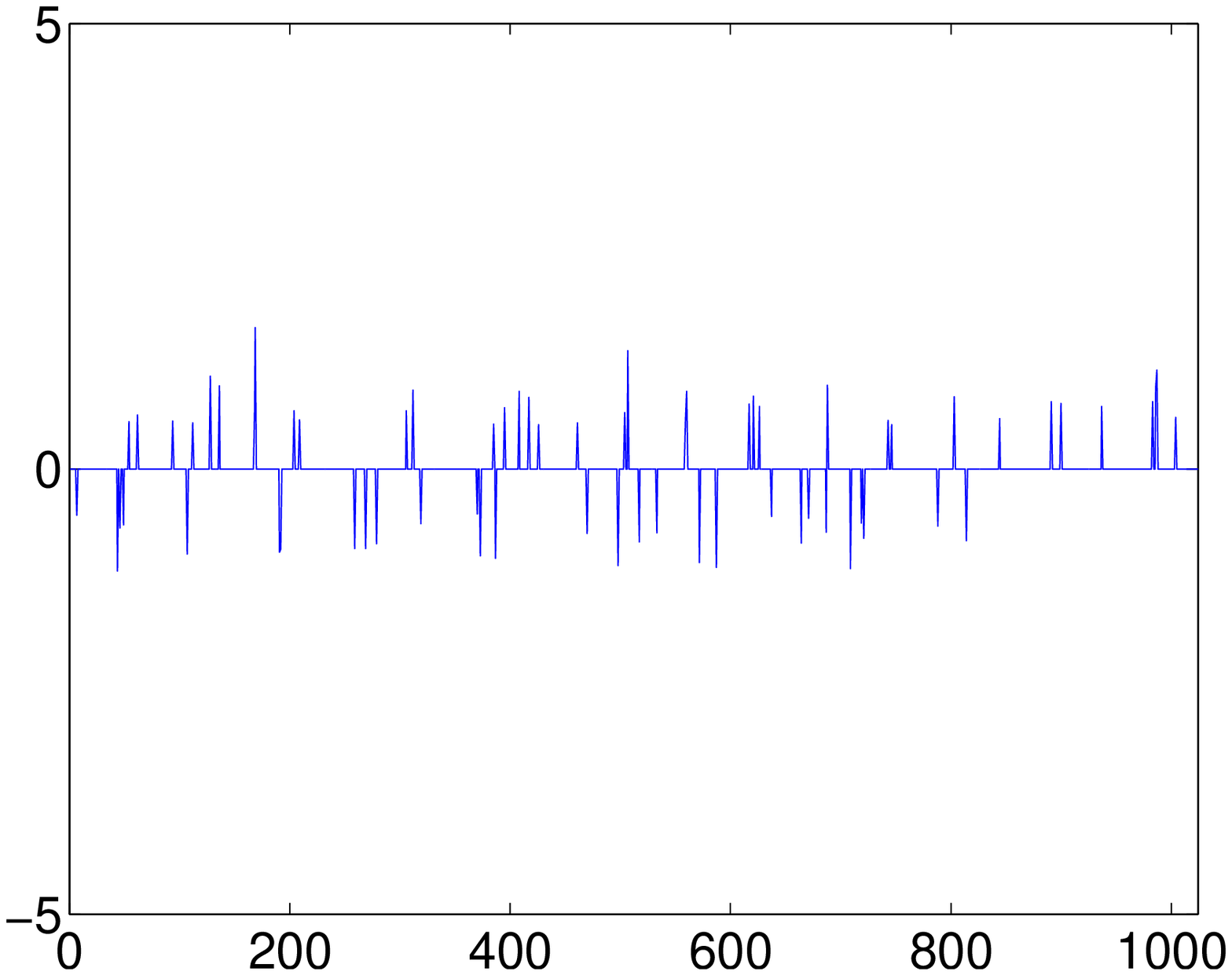}} &
{\includegraphics[width=0.3\hsize]{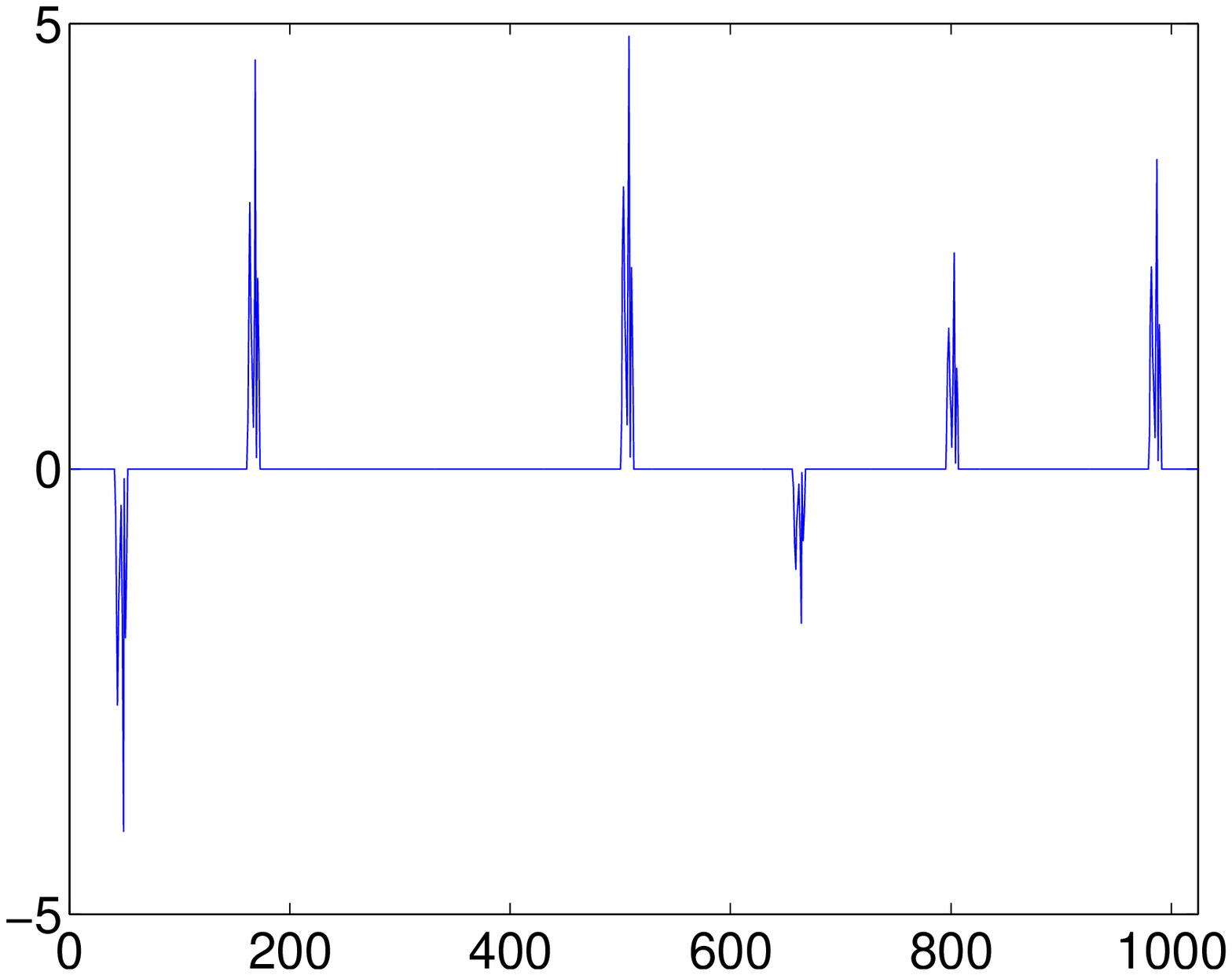}} \\
(c) & (d)
\end{tabular}
\caption{\small\sl 
(a) Test signal of length $N = 1024$ obtained by convolving a spike stream with $S = 6$ with an impulse response of length $F = 11$, so that the total signal sparsity $K =SF=66$. (b) Profile of one pulse . Signal reconstruction from $M=100$ random Gaussian measurements performed using (c) a state-of-the-art
CS recovery algorithm (CoSaMP~\cite{cosamp}, MSE = 13.42), and (d) our proposed Algorithm~\ref{alg:csdecon2} (MSE = 0.0028). 
\label{fig:ex1}}
\end{figure}

In this paper, we take some initial steps towards such a CS pulse stream framework. First, we propose a deterministic signal model for pulse streams. 
We show that our proposed model is equivalent to an {\em infinite union of subspaces}.
Second, as our main theoretical contribution, we derive a bound on the number of random linear measurements $M$ required to preserve the essential information contained in such signals. The proof relies on the particular high-dimensional geometry exhibited by the proposed model. Our derivation shows that $M = \bigo{(S + F) \log N}$; i.e., $M$ is proportional to the number of degrees of freedom of the signal $S + F$ but {\em sublinear} in the total sparsity $K = SF$.
Third, we develop algorithms to recover signals from our model from $M$ measurements. Under certain additional restrictions on the signals of interest, one of the algorithms provably recovers both the spike stream and the impulse response. We analyze its convergence, computational complexity, and robustness to variations in the pulse shape.  Numerical experiments on real and synthetic data sets demonstrate the benefits of the approach. As demonstrated in Figure~\ref{fig:ex1}, we obtain significant gains over conventional CS recovery methods, particularly in terms of reducing the number of measurements required for recovery.

This paper is organized as follows. In Section~\ref{sec:back}, we review the rudiments of standard and structured sparsity-based CS. In Section~\ref{sec:fss}, we propose a deterministic signal model for pulse streams and discuss its geometric properties. In Section~\ref{sec:rip}, we derive bounds on the number of random measurements required to sample signals belonging to our proposed model. In Section~\ref{sec:rec}, we develop an algorithm for stable signal recovery and analyze its convergence and robustness to model mismatch. Numerical results are presented in Section~\ref{sec:exp}, followed by a concluding discussion in Section~\ref{sec:conc}.
\section{Background on Compressive Sensing}
\label{sec:back}
\subsection{Sparse signal models}

A signal $x \in \real^N$ is $K$-{\em sparse} in the orthonormal basis $\Psi \in \real^{N \times N}$ if the corresponding basis representation $\alpha = \Psi^T x$ contains no more than $K$ nonzero elements. 
Without loss of generality, we assume the sparsity basis $\Psi$ to be the identity matrix for $\real^N$. 
The locations of the nonzeros of $x$ can additionally be encoded by a binary vector of length $N$ with a 1 indicating a nonzero; this vector $\sigma(x)$ is called the {\em support} of $x$. Denote the set of all $K$-sparse signals in $\real^N$ as $\Sigma_K$. Geometrically, $\Sigma_K$ can be identified as the union of ${N \choose K}$ subspaces of $\real^N$, with each subspace being the linear span of exactly $K$ canonical unit vectors of $\real^N$. For a general $x \in \real^N$, we define its best $K$-sparse approximation $x_K$ as
$$
x_K = \arg \min_{u \in \Sigma_K} \| x -u \|_2 .
$$

Many signals of interest exhibit more complex dependencies in terms of their nonzero values and locations. For instance, 
signals that permit only a small number of admissible support configurations can be modeled by a restricted union of subspaces, consisting only of $L_K$ canonical subspaces (so that $L_K$ is typically much smaller than ${N \choose K}$).  Thus, if $\Sigma =  \{ \sigma_1, \sigma_2, \ldots,  \sigma_{L_K} \}$ denotes the restricted set of admissible supports, then a {\em structured sparsity model}~\cite{modelCS} is the set
\begin{equation}
\A_K := \{ x : \sigma(x) \in \Sigma \} .\label{eq:ssm}
\end{equation}

\subsection{Signal acquisition via nonadaptive linear measurements}

Suppose that instead of collecting all the coefficients of a vector $x \in \real^N$, we merely record $M$ inner products (measurements) of $x$ with $M < N$ pre-selected vectors; this can be represented in terms of a linear transformation $y = \Phi x, \Phi \in \real^{M \times N}$. $\Phi$ is called the {\em sampling matrix}; it is at most rank-$M$ and hence has a nontrivial nullspace. The central result in Compressive Sensing (CS) is that despite the non-invertible nature of $\Phi$, if $x$ is {\em sparse}, then it can be exactly recovered from $y$ if $\Phi$ satisfies a condition known as the restricted isometry property (RIP):
\begin{DEFI}~\cite{CandesCS}
An $M\times N$ matrix $\Phi$ has the $K$-RIP with constant $\delta_K$ if, for all
$\x \in \Sigma_K$,
\begin{equation}
(1-\delta_K) \|\x\|_2^2 \le \|\Phi\x\|_2^2 \le (1+\delta_K)
\|\x\|_2^2. \label{eq:rip}
\end{equation}
\end{DEFI}

A matrix $\Phi$ with the $K$-RIP essentially ensures a {\em stable embedding} of the set of {\em all} $K$-sparse signals $\Sigma_K$ into a subspace of dimension $M$. The RIP requires $\Phi$ to leave the norm of every sparse signal approximately invariant; also, $\Phi$ must necessarily not contain any sparse vectors in its nullspace. 
At first glance, it is unclear if a matrix $\Phi$ that satisfies the RIP should even exist if $M < N$; indeed, deterministic design of a sampling matrix having the RIP is an NP-complete problem. Nevertheless, it has been shown~\cite{CandesCS} that provided $M \ge \bigo{K \log(N/K)}$,  
a matrix $\Phi$ whose elements are i.i.d. samples from a random subgaussian distribution possesses the RIP with high probability. Thus, $M$ can be linear in the sparsity of the signal set $K$ and {\em only logarithmic} in the signal length $N$. 

An analogous isometry condition holds for structured sparsity models containing $L_K$ canonical subspaces~\cite{samplingunion,modelCS,EldarUSS}. This is known as the {\em model-based RIP} and is defined thus: $\Phi$ satisfies the {\em $\A_K$-RIP} if (\ref{eq:rip}) holds for all $\x \in \A_K$.  It can be shown~\cite{samplingunion} that the number of measurements $M$ necessary for a subgaussian sampling matrix to have the $\A_K$-RIP with constant $\delta$ and with probability $1-e^{-t}$ is bounded as
\begin{equation}
M \ge \frac{c}{\delta^2}\left(\ln(2 L_K) + K \ln \frac{12}{\delta}+t \right).
\label{eq:blum}
\end{equation}
We can make two inferences from (\ref{eq:blum}). First, the number of measurements $M$ is logarithmic in the {\em number} of subspaces in the model; thus, signals belonging to a more concise model can be sampled using fewer random linear measurements. Second, $M$ is {\em at least linear} in the sparsity $K$ of the measured signal.

\subsection{Recovery methods}

Given measurements $y = \Phi x$, CS recovery methods aim to find the ``true'' sparse signal $x$ that generated $y$. One possible method is to seek the sparsest $x$ that generates the measurements $y$, i.e.,
\begin{equation}
\widehat x = \arg \min_{x'} \| x' \|_0~~\textrm{subject to}~~y = \Phi x' .
\label{eq:l0}
\end{equation}
where the $\ell_0$ ``norm'' of a vector $x'$ denotes the number of nonzero entries in $x'$.  This method can be used to obtain the true solution $x$ provided $M \geq 2K$. However, minimizing the $\ell_0$ norm can be shown to be NP-complete and is not stable in the presence of noise in the measurements~\cite{CandesCS}. 

If the sampling matrix $\Phi$ possesses the RIP, then tractable algorithms for CS recovery can be developed. These broadly follow two different approaches. The first approach entails solving a convex relaxation of (\ref{eq:l0}), e.g.,
\begin{equation}
\widehat x = \arg \min_{x'} \| x' \|_1~~\textrm{subject to}~~y = \Phi x' ,
\label{eq:l1}
\end{equation}
which corresponds to a linear program and hence can be solved in polynomial time. A common variant of this formulation includes accounting for noise of bounded magnitude in the measurements~\cite{BPDN}. The second approach entails an iterative, greedy selection of the support $\sigma(x)$ of the true solution $x$. This approach is employed by several algorithms such as orthogonal matching pursuit (OMP)~\cite{OMP}, compressive sampling matching pursuit (CoSaMP)~\cite{cosamp}, and iterative hard thresholding~\cite{iht}.

Both kinds of approaches provide powerful stability guarantees in the presence of noise while remaining computationally efficient. Given noisy measurements of any signal $x \in \real^N$ so that $y = \Phi x + n$, if $\Phi$ possesses the RIP, then the signal estimate $\widehat x$ obtained by these algorithms has bounded error:
\begin{equation}
\| x - \xhat \|_2 \leq C_1 \|x-x_K\|_2 + \frac{C_2}{\sqrt{K}}
\|x-x_K\|_1 + C_3 \|n\|_2,
\label{eq:recbound}
\end{equation}
where $x_K$ is the best $K$-sparse approximation to $x$ and $C_1, C_2$ are constants.  Furthermore, with a simple modification, algorithms like CoSaMP and iterative hard thresholding can be used to reconstruct signals belonging to any structured sparsity model~\cite{modelCS}.

To summarize, at the core of CS lie three key concepts: a signal model exhibiting a particular type of low-dimensional geometry in high-dimensional space, a low-rank linear mapping that provides a stable embedding of the signal model into a lower dimensional space, and algorithms that perform stable, efficient inversion of this mapping. 


\section{Signal Models for Pulse Streams}
\label{sec:fss}
Our objective is to extend the CS theory and algorithms to pulse stream signals. The conventional sparse signal model $\Sigma_K$ does not take into account the dependencies between the values and locations of the nonzeros in such signals. Indeed, these dependencies cannot be precisely captured by any structured sparsity model $\A_K$ that merely comprises a reduced subset of the subspaces in $\Sigma_K$. This necessitates richer models that capture the {\em convolutional} structure present in the nonzero coefficients of pulse streams. 

\subsection{General model}

Consider the following deterministic model for signals that can modeled by the convolution of an $S$-sparse spike stream $x \in \real^N$ with an $F$-sparse impulse response $h \in \real^N$. 
\begin{DEFI}
Let $\A_S \subset \real^N$ be a union of $S$-dimensional canonical subspaces, as defined in (\ref{eq:ssm}). Similarly, let $\A_F \subset \real^N$ be a union of $F$-dimensional canonical subspaces. Consider the set
\begin{equation}
\A^z_{S,F} := \{ z \in \real^N : z = x \ast h,~\textrm{such that}~x \in \A_S~\textrm{and}~h \in \A_F\} ,
\label{eq:fss}
\end{equation}
where $\ast$ denotes the circular convolution operator. Then, $\A^z_{S,F}$ is called a {\em pulse stream model}.
\label{def:psm}
\end{DEFI}

We make two immediate observations:
\subsubsection{Commutativity} Owing to the commutative property of the convolution operator, an element $z$ in $\A^z_{S,F}$ can be represented in multiple ways:
\begin{equation}
z=x \ast h = h \ast x = Hx = Xh ,
\label{eq:filt}
\end{equation}
where $H$ (respectively, $X$) is a square circulant matrix with its columns comprising circularly shifted versions of the vector $h$ (respectively, $x$). Therefore, Definition~\ref{def:psm} remains unchanged if the roles of $x$ and $h$ are reversed. We exploit this property during signal recovery from CS measurements in Section~\ref{sec:rec}.
\subsubsection{Geometry} It is useful to adopt the following geometric point of view: for a fixed $h \in \A_F$, the set $\{h \ast x: x \in \A_S\}$ forms a finite union of $S$-dimensional subspaces, owing to the fact that it is generated by the action of $h$ on $L_S$ canonical subspaces. Denote this set by $h(\A_S)$. Then, the pulse stream model in (\ref{eq:fss}) can be written as
$$
\A^z_{S,F} = \bigcup_{h \in \A_F} h(\A_S) .
$$
Thus, our signal model can be interpreted as an {\em infinite union of subspaces}.\footnote{A general theory for sampling signals from infinite unions of subspaces has been introduced in~\cite{dosamplingunion}.} Note that (\ref{eq:blum}) cannot be applied in this case since it only considers finite unions of subspaces. However, let $K =SF$ denote the maximum sparsity of the signals in Definition~\ref{def:psm}. Then, it is clear that the set $\A^z_{S,F}$ is a very small subset of $\Sigma_K$, the set of all $SF$-sparse signals. We exploit this property while proving our main sampling results in Section~\ref{sec:rip}. 

Note that the exact definition of convolution operator changes depending on the domain of the signals of interest. For one-dimensional (1D) time domain signals of length $N$, the square matrix $H$ is formed by all $N$ circular shifts of the vector $h$; for 2D images of size $N$ pixels, $H$ is formed by all 2D circular shifts of $h$, and so forth.

\subsection{Special case: Disjoint pulses}

The model proposed in Definition~\ref{def:psm} is general and applicable even to signals in which successive pulses overlap with each other. In Section~\ref{sec:rip} we develop a lower bound on the number of samples required to preserve the essential information contained in an arbitrary pulse stream. However, feasible recovery of such general pulse streams from CS measurements is rather difficult; we examine this in detail in Section~\ref{sec:rec}. Therefore, we will also consider a more restrictive model where the pulses are assumed to not overlap.

For concreteness, consider 1D time domain signals as specified by
(\ref{eq:filt}). Note that $H$ and $x$ need not be unique for a given $z$;  any ordered pair $(\alpha H, x / \alpha)$ satisfies (\ref{eq:filt}), and so does $(H',x')$, where $H'$ is generated by a circularly shifted version of $h$ by a time delay $+\tau$ and $x'$ is a circularly shifted version of $x$ by $-\tau$. To eliminate these ambiguities, we make the following two assumptions:
\begin{enumerate}
\item the impulse response $h$ is {\em concentrated}, i.e., all the nonzero coefficients of $h$ are contiguously located in its first $F$ indices. Thus, the structured sparsity model $\A_F$ for the vector $h$ consists of the lone subspace spanned by the first $F$ canonical unit vectors. 
\item the spikes are sufficiently separated in time. In particular, any two consecutive spikes in the vector $x$ are separated at least by $\Delta$ locations, where $\Delta \geq F$.  
A structured sparsity model for such time-domain signals with sufficiently separated nonzeros has been introduced in~\cite{spikemodel}. 
\end{enumerate}

The notion of disjoint pulses can be immediately generalized to signals defined over domains of arbitrary dimension. Consider $S$-sparse spike streams $x$ defined over a domain of dimension $n$. Suppose that  at most one spike in $x$ can occur in a hypercube in $\real^n$ with side $\Delta$. This defines a special structured sparsity model for the spike streams of interest; denote this model as $\A_S^\Delta$. Further, let the $F$ nonzero coefficients in $h$ be concentrated within a hypercube centered at the domain origin whose side length is no greater than $\Delta$. Then, a deterministic model for sums of non-overlapping pulses of arbitrary dimension can be proposed as follows.
\begin{DEFI}
Let $\A^\Delta_S$ be the structured sparsity model for spike streams as defined above. Let $\A_F$ be the subspace of concentrated impulse responses of sparsity $F$. Define the set
\begin{equation}
\A(S,F, \Delta) = \{ z \in \real^N : z = x \ast h,\text{such that}~x \in \A^\Delta_S~\text{and}~h \in \A_F\} .
\end{equation}
Then, $\A(S,F,\Delta)$ is called the {\em disjoint pulse stream model}.
\end{DEFI}

This model eliminates possible ambiguities that arise due to the shift-invariant nature of convolution; i.e., the locations of the nonzero spikes that generate a disjoint pulse stream are {\em uniquely} defined. 
This property proves to be essential in developing and analyzing a feasible method for signal recovery (Section~\ref{sec:rec}). See Figure~\ref{fig:ex1}(a) for an example stream of disjoint pulses in 1D.

\section{Sampling Theorems for Pulse Streams}
\label{sec:rip}
Pulse streams can be modeled as an infinite union of low-dimensional subspaces. The next ingredient in the development of a CS framework for such signals is a bound on the number of linear samples required to preserve the essential information of this signal set.

\subsection{General pulse streams}

We derive a sampling theorem for signals belonging to the model  $\A^z_{S,F}$ proposed in Definition~\ref{def:psm}. Suppose that $K=SF$. As mentioned above, $\A^z_{S,F}$ is a subset of the set of all $K$-sparse signals $\Sigma_{K}$. On the other hand, only a small fraction of all $K$-sparse signals can be written as the convolution of an $S$-sparse spike stream with an $F$-sparse impulse response. Thus, intuition suggests that we should be able to compressively sample signals from this set using fewer random linear measurements than that required for the set of all $K$-sparse signals. The following theorem makes this precise.
\begin{THEO}
Suppose $\A^z_{S,F}$ is the pulse stream model from Definition~\ref{def:psm}. Let $t > 0$. Choose an $M \times N$ i.i.d. subgaussian matrix $\Phi$ with
\begin{equation}
M \ge \bigo{\frac{1}{\delta} \left( (S + F)\ln\left(\frac{1}{\delta}\right) +  \log(L_S L_F) + t \right) }.
\label{eq:sfrip}
\end{equation}
Then, $\Phi$ satisfies the following property with probability at least $1 - e^{-t}$: for every pair $z_1, z_2 \in \A^z_{S,F}$,
\begin{equation}
(1-\delta) \|\z_1 - z_2\|_2^2 \le \|\Phi z_1 - \Phi z_2\|_2^2 \le (1+\delta)
\|z_1 - z_2\|_2^2. \label{eq:zrip}
\end{equation}
\label{thm:zrip}
\end{THEO}

The proof of this theorem is presented in Appendix~\ref{app:jl}.  
An important consequence of the theorem is that, by definition, $\A_S$ is a subset of the set of all $S$-dimensional canonical subspaces. In particular,
\begin{equation}
L_S \leq {N \choose S} \approx \left( \frac{eN}{S} \right)^S .
\end{equation}
Similarly, $L_F \leq  \left( \frac{eN}{F} \right)^F .$
Therefore, the logarithmic term in the expression for $M$ in (\ref{eq:sfrip}) scales as:
\begin{equation}
\log(L_S L_F) \leq  S + S \log(N/S) + F + F \log(N/F) \leq 2(S + F) \log N 
\end{equation}
Thus, (\ref{eq:sfrip}) indicates that the number of measurements $M$ required for the sampling of signals in $\A^z_{S,F}$ is proportional to $(S + F)$. Therefore, $M$ is {\em sublinear}
 in the total sparsity of the signals $K = SF$. In contrast, conventional structured sparsity models would require at least $2K = 2SF$ linear measurements to ensure a stable embedding of the signal set~\cite{samplingunion}. In addition, the number of degrees of freedom of each signal can be considered to be $\bigo{S + F}$, corresponding to the positions and locations of the coefficients of the sparse signal and impulse response. Therefore, the bound in Theorem~\ref{thm:zrip} is essentially optimal for the signal class $\A^z_{S,F}$.

\subsection{Special case: Disjoint pulse streams}

Theorem~\ref{thm:zrip} is valid for signals belonging to the general model $\A_{S,F}^z$. In the case of disjoint pulse streams, we can derive a more stringent lower bound. By definition, the $F$ nonzero coefficients of $h$ are concentrated in a hypercube around the domain origin. Therefore, $h$ lies in a lone subspace spanned by $F$ basis vectors of $\real^N$, and hence $L_F = 1$. Further, a simple modification of Theorem 1 of~\cite{spikemodel} states that the number of subspaces in the structured sparsity model $\A_S^\Delta$ is given by
\begin{equation}
L_S = \binom{N - S\Delta + S-1}{S-1} .
\label{eq:suppcount}
\end{equation}
Thus, for the disjoint pulse stream model $\A(S,F,\Delta)$, we obtain the following easy corollary to Theorem~\ref{thm:zrip}. 
\begin{COROLLARY}
If $t > 0$ and 
\begin{equation}
M \ge \bigo{\frac{1}{\delta} \left( (S + F)\ln\left(\frac{1}{\delta}\right) +  S \log(N/S - \Delta) + t \right) },
\label{eq:sfdrip}
\end{equation}
then an $M \times N$ i.i.d.\ gaussian matrix $\Phi$ will satisfy (\ref{eq:zrip}) with probability at least $1 - e^{-t}$ for any pair of signals $z_1, z_2$ belonging to the $\A(S,F,\Delta)$ model.
\label{corr:deltam}
\end{COROLLARY}

Note that the parameter $\Delta$ can be at most $N/S$, since $S$ spikes must be packed into $N$ coefficient locations with at least $\Delta$ locations separating any pair of spikes. A higher value of $\Delta$ implies that the model $\A^\Delta_S$ admits a smaller number of configurations; thus, (\ref{eq:sfdrip}) implies that fewer measurements are needed to sample pulse streams in which the pulses are widely separated. 

\section{Recovery of Pulse Streams}
\label{sec:rec}
The final ingredient in our extended CS framework for pulse streams consists of new algorithms for the stable recovery of the signals of interest from compressive measurements. This problem can be stated as follows. Suppose $z \in \A_{S,F}^\z$. If w are given the noisy measurements
$$
y = \Phi z + n = \Phi H x + n = \Phi X h + n , 
$$
then we aim to reconstruct $z$ from $y$.  The main challenge stems from the fact that {\em both} $x$ (respectively, $X$) and $h$ (respectively, $H$) are unknown and have to be simultaneously inferred. 

This problem is similar to performing sparse approximation with {\em incomplete} knowledge of the dictionary in which the target vector (either $x$ or $h$) is sparse.
This problem has received some interest in the literature~\cite{bestbasis,herobd,cspt}; the common approach has been to first assume that a training set of vectors $\{x_i\}$ exists for a fixed impulse response $h$, then infer the coefficients of $h$ using a sparse learning algorithm (such as LASSO~\cite{lasso} or basis pursuit~\cite{BPDN}), and then solve for the coefficients $\{x_i\}$. 
In the absence of training data, we must infer both the spike locations and the impulse response coefficients.  Therefore, our task is also similar to {\em blind deconvolution}~\cite{bdecon}; the main differences are that we are only given access to the random linear measurements $y$ as opposed to the Nyquist rate samples $z$, and that our primary aim is to reconstruct $z$ as faithfully as possible as opposed to merely reconstructing $x$.

Our general approach will be to fix an estimate of $h$, obtain the ``best possible" estimate of $x$, update our estimate of $h$, and iterate. This is commonly known as {\em alternating minimization} (AM) and has been shown to be suitable for blind deconvolution settings~\cite{tchan}. As demonstrated below in the proof of Theorem~\ref{thm:alg1}, we require that the best possible estimate of the spike stream $x$ and the impulse response $h$ at each iteration are unique. For this reason, we will assume that our target signal $z$ belongs to the disjoint pulse stream model $\A(S,F,\Delta)$.

\subsection{Alternating minimization with exhaustive search}
\label{subsec:am}
Consider $z \in \A(S,F,\Delta)$, so that $z = x \ast h$.
This implies that the spikes in $x$ are separated by a minimum separation distance $\Delta$ and that the impulse response $h$ is concentrated. Suppose first that we are given noiseless CS measurements $y = \Phi z$.
We fix a candidate support configuration $\sigma$ for the spike stream (so that $\sigma$ contains $S$ nonzeros.) Then, we form the circulant matrix $\widehat H$ from all possible shifts of the current estimate of the impulse response $\hhat$ (denote this operation as $\widehat H=\mathbb{C}(\hhat)$). Further, we calculate the dictionary $\Phi \widehat H$ for the spike stream $x$, and select the submatrix formed by the columns indexed by the assumed spike locations $\sigma$ (denote this submatrix as $(\Phi \widehat H)_\sigma$). This transforms our problem into an overdetermined system, which can be solved using least-squares. 
In summary, we use a simple matrix pseudoinverse to obtain an estimate for $\widehat x$:
$$
\widehat x = (\Phi \widehat{H})_{\sigma}\pinv y .
$$
This gives us an estimate of the spike coefficients $\xhat$ for the assumed support configuration $\sigma$.  We now exploit the commutativity of the convolution operator $\ast$. We form the circulant matrix $\widehat X$, form the dictionary $\Phi \widehat X$ for the impulse response and select the submatrix $(\Phi \widehat{X})_f$ formed by its first $F$ columns. Then, we solve a least-squares problem to obtain an estimate $\hhat$ for the impulse response coefficients: 
$$
\widehat h = (\Phi \widehat{X})_f\pinv y .
$$
Then, we form our signal estimate $\widehat z = \xhat \ast \hhat$. The above two-step process is iterated until a suitable halting criterion (e.g., convergence in norm for the estimated signal $\widehat z$). This process is akin to the Richardson-Lucy algorithm for blind deconvolution~\cite{starck}. 

The overall reconstruction problem can be solved by repeating this process for every support configuration $\sigma$ belonging to the structured sparsity model $\A_S^\Delta$ and picking the solution with the smallest norm of the residual $r = y - \Phi \widehat z$.
The procedure is detailed in pseudocode form in Algorithm~\ref{alg:csdecon1}.
\begin{algorithm*}[!t]
\caption{Alternating minimization with exhaustive search} 
\label{alg:csdecon1}
\begin{tabbing}
Inputs: Sampling matrix $\Phi$, measurements $\y = \Phi x$, model parameters $\Delta$, $S$, $F$, threshold $\epsilon$ \\
Output: $\widehat z \in \A(S,F,\Delta)$ such that $y - \Phi \widehat z$ is small  \\
$\xhat=0$, $\widehat h = (\mathbf{1}^T_F, 0, \ldots, 0) / \sqrt{F}$; $i = 0$ \hspace{18mm} \{initialize\} \\
{\bf for} \= $\sigma \in \A_S^\Delta$ {\bf do} \hspace{55mm}\= \\
\> 1. $\widehat H = \mathbb{C}(\hhat), \Phi_h = (\Phi \widehat H)_\sigma$ \> \{form dictionary for spike stream\} \\
\> 2. $\xhat \leftarrow \Phi_h\pinv y$ \> \{update spike stream estimate \} \\
\> 3. $\widehat X = \mathbb{C}(\xhat), \Phi_x = (\Phi \widehat X)_f$ \> \{form dictionary for impulse response\} \\
\> 4. $\hhat \leftarrow \Phi_x\pinv y$ \> \{update impulse response estimate\} \\
\> 5. $\zhat \leftarrow \xhat \ast \hhat$ \> \{form signal estimate\} \\
{\bf if} \= $\|y- \Phi \zhat\|_2 <  \epsilon$ \> \{check for energy in residual\}  \\
\>  {\bf return} $\zhat$\\
{\bf end if} \\
{\bf end for}
\end{tabbing}
\vspace*{-3mm}
\end{algorithm*}
Thus, Algorithm~\ref{alg:csdecon1} consists of performing alternating minimization for a given estimate for the support of the underlying spike stream $x$, and exhaustively searching for the best possible support.  Under certain conditions on the sampling matrix $\Phi$, we can study the convergence of Algorithm~\ref{alg:csdecon1} to the correct answer $z$, as encapsulated in the following theorem.
\begin{THEO}
Let $z \in \A(S,F,\Delta)$ and suppose that $\Phi$ satisfies (\ref{eq:zrip}) with constant $\delta$ for signals belonging to $\A(S,F,\Delta)$. Suppose we observe $y = \Phi z$ and apply Algorithm~\ref{alg:csdecon1} to reconstruct the signal. Let $\widehat z_i$ be an intermediate estimate of $z$ at iteration $i$ of Algorithm~\ref{alg:csdecon1}. Then:
\begin{enumerate}
\item The norm of the residual $\|y - \Phi \widehat z_i \|_2$ monotonically decreases with iteration count $i$.
\item If at any iteration $i$
$$
\|y - \Phi \widehat z_i \|_2 \leq \epsilon,
$$
then we are guaranteed that
$$
\|z - \widehat z_i\|_2 \leq c \epsilon,
$$
where $c$ depends only on $\delta$.
\end{enumerate}
\label{thm:alg1}
\end{THEO}
The proof of this theorem is provided in Appendix~\ref{app:alg}. The first part of the theorem implies that for any given support configuration $\sigma$, Steps 1 through 4 in Algorithm~\ref{alg:csdecon1} are guaranteed to converge to a generalized fixed point~\cite{amtropp}. The second part of the theorem provides a condition on the detection of the true support configuration $\sigma$ 
in the following weak sense: if the energy of the residual of the signal estimate is small, then the signal has been accurately reconstructed.

\subsection{Model mismatch}
\label{subsec:mm}

In practical situations, we would expect to have minor variations in the shapes of the $S$ pulses in the signal $z$. In this case, $z$ can no longer be expressed as $H x$ where $H$ is a circulant matrix. Let $\{h_1, h_2, \ldots, h_S\}$ be length-$F$ vectors corresponding to each of the $S$ pulses in the signal, and let the length-$S$ spike stream $\widetilde{x} = (\alpha_1, \alpha_2, \ldots, \alpha_S)$. Further, let $\mathbb{S}_i$ be the circular shift operator that maps the $i^{\textrm{th}}$ pulse shape $h_i$ into its corresponding vector in $\real^N$. Then, we have
\begin{equation}
z = \sum_{i = 1}^S \alpha_i \mathbb{S}_i(h_i),
\label{eq:varpuls}
\end{equation}
or equivalently, 
$$
z = \widetilde{H} \widetilde{x},
$$
where $\widetilde{H} = [\mathbb{S}_1(h_1), \ldots, \mathbb{S}_S (h_S)]$ is an $N \times S$ matrix. 
Assuming that the spikes in $x$ are separated by at least $\Delta$ locations, the matrix $\widetilde{H}$ is {\em quasi-Toeplitz}~\cite{cri}, i.e., the columns of $\widetilde{H}$ are circular shifts of one another with no more than one nonzero entry in every row. An attractive property of quasi-Toeplitz matrices is that there exist analytical expressions for their pseudo-inverses. 
 Suppose the measurement matrix $\Phi$ equals the identity, i.e., we are given Nyquist-rate samples of $z$. Then the matrices $\Phi_h$ and $\Phi_x$ in Step 2 of Algorithm~\ref{alg:csdecon1} are also quasi-Toeplitz, and hence $\Phi_h\pinv$ and $\Phi_x\pinv$ can be computed in closed form.  Thus, given an estimate of the pulse shape $\hhat_0$, we can derive closed-form expressions for the next impulse reponse estimate. 

Additionally, we can obtain an intermediate estimate for the spike stream $\widetilde{x}$. Suppose the innermost loop of Algorithm~\ref{alg:csdecon1} converges to a fixed point estimate $\hhat$. Since the least-squares equations are homogenous, we may assume that $\| \hhat \|_2 = 1$ without loss of generality. We dub $\hhat$ the {\em anchor pulse} for the set of pulse shapes $\{h_1, h_2, \ldots, h_S\}$. The following theorem provides an expression relating the anchor pulse to the component pulse shapes.

\begin{THEO}
Consider $z$ as defined in (\ref{eq:varpuls}). Let $\hhat$ be the anchor pulse for the set of pulse shapes $\{h_1, h_2, \ldots, h_S\}$. Define $c_i = \langle h_i, \hhat \rangle$ for $i = 1,\ldots,S$.~
Then, we have that
\begin{equation}
\hhat = \frac{\sum_{i=1}^S c_i \alpha_i^2 h_i}{\sum_{i=1}^S c_i^2 \alpha_i^2} .
\label{eq:ap}
\end{equation}
\label{thm:anchor}
\end{THEO}
The proof of this theorem is provided in Appendix~\ref{app:anchor}. Equation~(\ref{eq:ap}) implies that the anchor pulse $\hhat$ is a weighted linear combination of the component pulses $h_i,~i = 1,\ldots,S$, with the weights defined by the corresponding spike coefficients $\alpha_i$ and the inner products $c_i$. 
The anchor pulse remains unchanged if the spike coefficient vector $\widetilde{x}$ is multiplied by any constant $C$. Therefore, the anchor pulse can be viewed as a {\em scale-invariant average} of the component pulse shapes.

Theorem~\ref{thm:anchor} applies to Nyquist-rate samples of the signal $z$. In the case of low-rate CS measurements $y = \Phi z$,  the convergence analysis of Algorithm~\ref{alg:csdecon1} for the general case of $S$ different pulse shapes becomes more delicate. If $\Phi$ possesses the RIP only for $z \in \A(S,F,\Delta)$, then it could be that two different pulse streams $z_1, z_2$ (each with varying shapes across pulses) are mapped by $\Phi$ to the same vector in $\real^M$, i.e., $\Phi z_1 = \Phi z_2$; thus, the unique mapping argument employed in the proof of Theorem~\ref{thm:alg1} cannot be applied in this case. 
One way to analyze this case is to recognize that by allowing arbitrary pulse shapes $\{h_1, h_2, \ldots, h_S\}$, our space of signals of interest is equivalent to a special structured sparsity model that consists of all $K$-sparse signals whose non-zeros are arranged in $S$ blocks of size $F$ and the starting locations of consecutive blocks are separated by at least $\Delta$ locations. As discussed in Section~\ref{sec:back}, stable CS reconstruction for signals from this model requires at least $M = 2SF = 2K$ measurements; thus, Algorithm~\ref{alg:csdecon1} converges in the general case given that $M$ is proportional to $K$. Thus, in the case of arbitrary pulse shapes, the number of measurements required by Algorithm~\ref{alg:csdecon1} is on the same order as the number of measurements required for conventional structured sparsity-based CS recovery.

\subsection{Iterative support estimation}
\label{subsec:fa}

Algorithm~\ref{alg:csdecon1} involves iteratively solving a combinatorial number of estimation problems. This becomes infeasible for even moderate values of $N$. A simpler method can be proposed as follows: instead of cycling through every possible support configuration $\sigma_i$ for the spike stream $x$, we instead retain an {\em estimate} of the support configuration, based on the current estimates of the spike stream $\xhat$ and impulse response $\hhat$, and update this estimate with each iteration.  In order to ensure that the support estimate belongs to $\A_S^\Delta$, we leverage a special CS recovery algorithm for signals belonging to $\A_S^\Delta$ that is based on CoSaMP~\cite{cosamp}. We provide an outline of the algorithm here for completeness; see~\cite{spikemodel} for details. 

At each iteration, given an estimate of the spike coefficients $x$, we need to solve for the best $\A^\Delta_S$-approximation to $x$. Let $x = (x_1, x_2, \ldots,x_N)^T$. Given any binary vector $s = (s_1, s_2, \ldots,s_N)^T$ of length $N$, let:
$$
x_{\vert s} := (s_1 x_1, s_2 x_2, \ldots, s_N x_N) ,
$$
so that $x_{\vert s}$ is the portion of the signal $x$ lying within the support $s$. Our goal is to solve for the choice of support $s$ so that $x_{\vert s}$ belongs to $\A^\Delta_S$ and $\|x - x_{\vert s} \|_2$ is minimized. The following constraints on the support vector $s$ follow from the definition of $\A_S^\Delta$:
\begin{eqnarray}
s_1 + s_2 + \ldots + s_N&\leq&S , \label{eq:sps}\\
s_j + s_{j+1} \ldots + s_{j + \Delta-1} &\leq&1,~\textrm{for}~j = 1,\ldots,N, \label{eq:ineq} 
\end{eqnarray}
where the subscripts are computed modulo $N$. The first inequality (\ref{eq:sps}) specifies that the solution contains at most $S$ nonzeros; the other $N$ inequalities (\ref{eq:ineq}) specify that there is at most one spike within any block of $\Delta$ consecutive coefficients in the solution.  

It can be shown that minimizing $\|x - x_{\vert s} \|_2$ is equivalent to maximizing $c^T s$ where $c = (x_1^2,x_2^2,\ldots,x_N^2)$, i.e., maximizing the portion of the energy of $x$ that lies within $s$. Define $W \in \real^{(N+1) \times N}$ as a binary indicator matrix that captures the left hand side of the inequality constraints  (\ref{eq:sps}) and (\ref{eq:ineq}). Next, define $u \in \real^{N+1} = (S,1,1,\ldots,1)$; this represents the right hand side of the constraints (\ref{eq:sps}) and (\ref{eq:ineq}).
Thus, we can represent (\ref{eq:sps}) and (\ref{eq:ineq}) by the following binary integer program:
\begin{equation*}
s^*=\arg \min~c^T s ,~\textrm{subject to}~W s \leq u. \\ 
\end{equation*} 
Next, we relax the integer constraints on $s$ to obtain a computationally tractable linear program. Denote this linear program by $\mathbb{D}(\cdot)$. In~\cite{spikemodel}, it is shown that the solutions to the integer program and its relaxed version are identical. Thus, we have a computationally feasible method to obtain an estimate of the support of the best  $\A^\Delta_S$-approximation to $x$. 

Once an updated support estimate has been obtained, we repeat Steps 2, 3 and 4 in Algorithm~\ref{alg:csdecon1} to solve for the spike stream $x$ and impulse $h$. This process is iterated until a suitable halting criterion (e.g., convergence in norm for the estimated pulse stream $\widehat z$.) The overall algorithm can be viewed as an iterative sparse approximation procedure for the $\A_S^\Delta$ model that continually updates its estimate of the sparsifying dictionary.  The procedure is detailed in pseudocode form in Algorithm~\ref{alg:csdecon2}.
\begin{algorithm*}[!t]
\caption{Iterative support estimation}
\label{alg:csdecon2}
\begin{tabbing}
Inputs: Sampling matrix $\Phi$, measurements $\y = \Phi z + n$, model parameters $\Delta$, $S$, $F$. \\
Output: $\A(S,F,\Delta)$-sparse approximation $\widehat z$ to true signal $z$ \\
Initialize $\xhat=0$ , $\widehat h = (\mathbf{1}^T_F, 0, \ldots, 0)$, $i = 0$ \hspace{29mm} \\
{\bf while} \= halting criterion false {\bf do} \hspace{30mm}\= \\
\> 1. $i \leftarrow i+1$ \\
\> 2. $\widehat z \leftarrow \xhat \ast \hhat$ \> \{current pulse stream estimate\}\\
\> \{estimate spike locations and amplitudes\} \\
\> 3. $\widehat H = \mathbb{C}(\hhat), \Phi_h = \Phi \widehat H$ \> \{form dictionary for spike stream\} \\
\> 4. $e \leftarrow \Phi_h^T (y - \Phi_h \xhat) $ \> \{residual\} \\
\> 5. $\omega \leftarrow \sigma(\mathbb{D}(e))$ \> \{obtain model-approximation of residual\} \\
\> 6. $\sigma \leftarrow \omega \cup \sigma(\xhat_{i-1})$ \> \{merge supports\} \\
\> 7. $x\vert_\sigma \leftarrow (\Phi_h)_\sigma\pinv y$, $x\vert_{\sigma^C} = 0$ \> \{update spike stream estimate\} \\
\> 8. $\xhat \leftarrow \mathbb{D}(x)$ \> \{prune spike stream estimate\} \\
\> \{estimate impulse response\} \\
\> 9. $\widehat X = \mathbb{C}(\xhat), \Phi_x = (\Phi \widehat X)_f $ \> \{form dictionary for impulse response\} \\
\> 10. $\hhat \leftarrow \Phi_x\pinv y$ \> \{update impulse response estimate\} \\
{\bf end while} \\
{\bf return} $\zhat \leftarrow \xhat \ast \hhat$
\end{tabbing}
\vspace*{-2mm}
\end{algorithm*}

\subsection{Stability and convergence}
Like many other algorithms for blind deconvolution, the analysis of Algorithm~\ref{alg:csdecon2} is not straightforward. The dictionaries $\Phi \widehat X$ and $\Phi \widehat H$ are only approximately known at any intermediate iteration, and hence the proof techniques employed for the analysis for CoSaMP do not apply. In principle, given access to a sufficient number of measurements, we may expect similar convergence behavior for Algorithm~\ref{alg:csdecon2} as Algorithm~\ref{alg:csdecon1}. Empirically, Algorithm~\ref{alg:csdecon2} can be shown to be stable to small amounts of noise in the signal as well as in the CS measurements and to minor variations in the pulse shape. We demonstrate this with the help of various numerical experiments in Section~\ref{sec:exp}. 

\subsection{Computational complexity}

The primary runtime cost of Algorithm~\ref{alg:csdecon2} is incurred in solving the linear program $\mathbb{D}(\cdot)$. For a length-$N$ signal, the computational complexity of solving a linear program is known to be $\bigo{N^{3.5}}$. The total computational cost also scales linearly in the number of measurements $M$ and the number of iterations $T$ of the outer loop executed until convergence; thus, overall the algorithm runs in polynomial time. 

\section{Numerical experiments}
\label{sec:exp}

We now present a number of results that validate the utility of our proposed theory and methods.
All numerical experiments reported in this section have been performed using Algorithm~\ref{alg:csdecon2} for recovery of disjoint pulse streams.

\subsection{Synthetic 1D pulse streams}
Figure~\ref{fig:ex1} demonstrates the considerable advantages that Algorithm~\ref{alg:csdecon2} can offer in terms of the number of compressive measurements required for reliable reconstruction. The test signal was generated by choosing $S = 8$ spikes with random amplitudes and locations and convolving this spike stream with a randomly chosen impulse response of length $F=11$. The overall sparsity of the signal $K = SF = 88$; thus, standard sparsity-based CS algorithms would require at least $2K = 176$ measurements. Our approach (Algorithm~\ref{alg:csdecon2}) returns an accurate estimate of both the spike stream as well as the impulse response using merely $M = 90$ measurements. 

\begin{figure}[!t]
\centering
{\includegraphics[width=0.4\hsize]{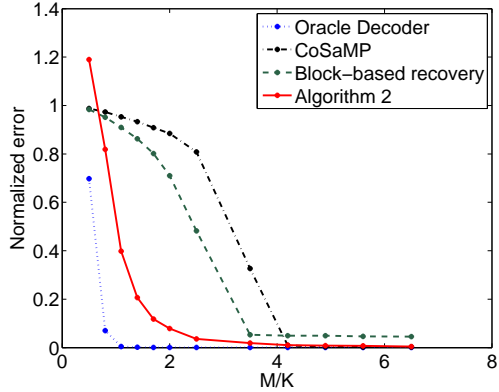}}
\caption{\small \sl  Normalized reconstruction MSE vs.\ $M/K$ for different reconstruction algorithms averaged over 200 sample trials. Signal parameters: $N = 1024$, $S = 8$, $F=11$.  Algorithm 2 outperforms standard and structured sparsity-based methods, particularly when $M/K$ is small.} \label{fig:err}
\end{figure}

Figure~\ref{fig:err} displays the averaged results of a Monte Carlo simulation of Algorithm~\ref{alg:csdecon2} over 200 trials. Each trial was conducted by generating a sample signal belonging to $\A(S,F,\Delta)$, computing $M$ linear random Gaussian measurements, reconstructing with different algorithms, and recording the magnitude of the recovery error for different values of $M/K$. It is clear from the figure that Algorithm~\ref{alg:csdecon2} outperforms both conventional CS recovery (CoSaMP~\cite{cosamp}) with target sparsity $K = SF$ as well as block-based reconstruction~\cite{modelCS} with knowledge of the size and number of blocks (respectively $F$ and $S$). In fact, our algorithm performs nearly as well as the ``oracle decoder'' that possesses perfect prior knowledge of the impulse response coefficients and aims to solve only for the spike stream.

\begin{figure}[!t]
\centering
\begin{tabular}{cc}
{\includegraphics[width=0.3\hsize]{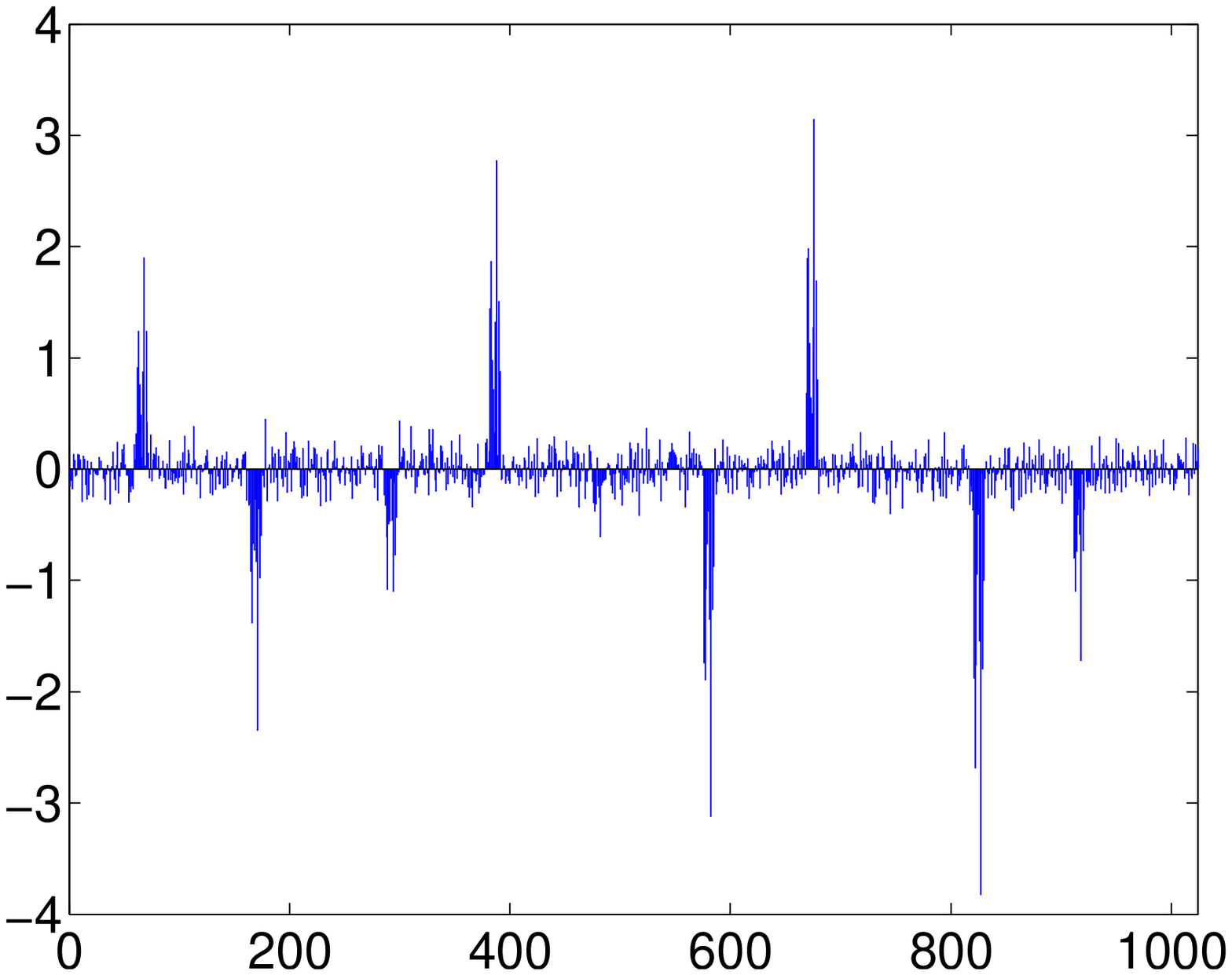}}&
{\includegraphics[width=0.3\hsize]{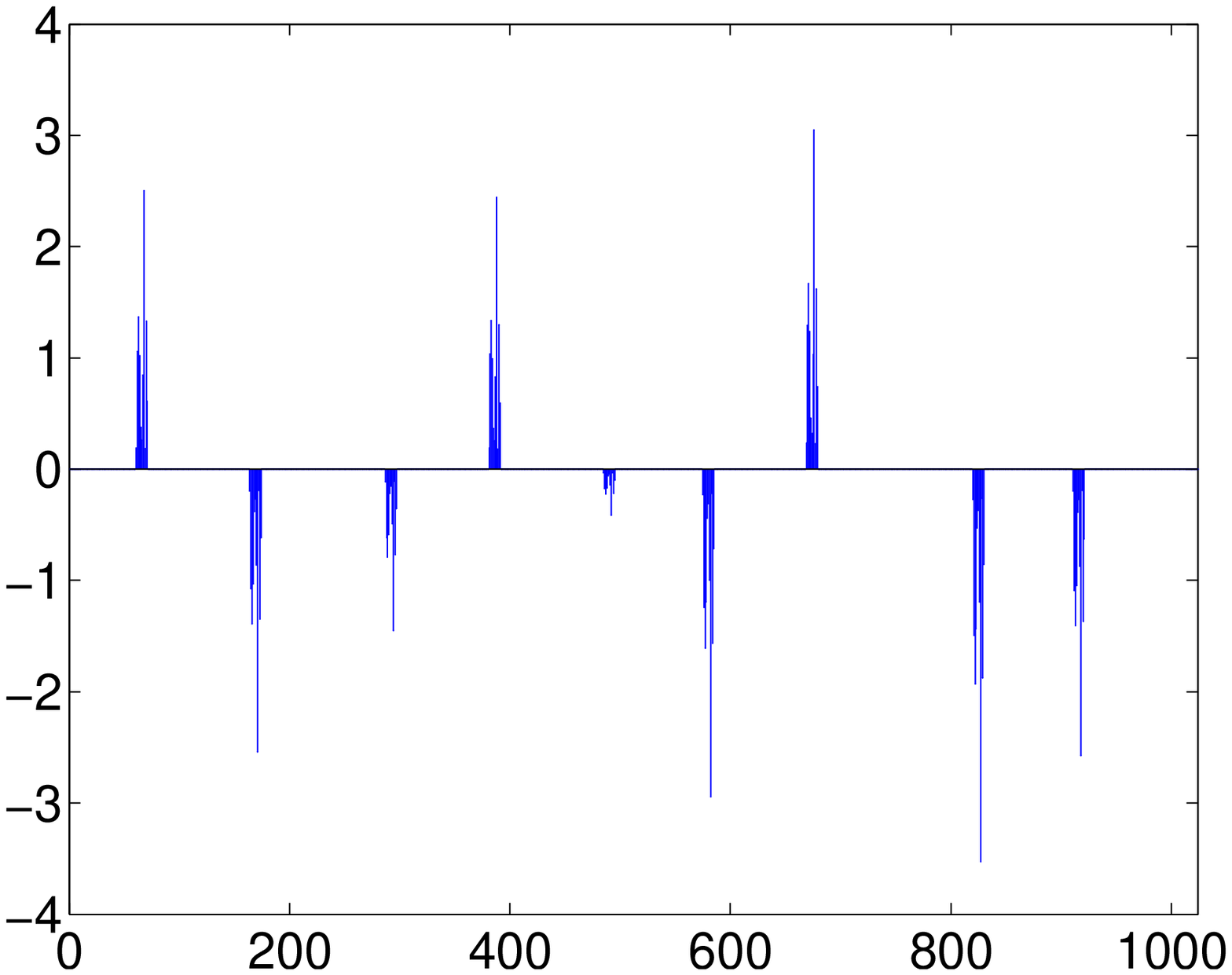}} \\
(a) & (b) 
\end{tabular}
\caption{\small\sl (a) Synthetic noisy signal (SNR = 13.25dB). (b) Recovery from $M=150$ random measurements using Algorithm~\ref{alg:csdecon2}. 
\label{fig:ex6}}
\end{figure}

We show that Algorithm~\ref{alg:csdecon2} is stable to small amounts of noise in the signal and the measurements. In Figure~\ref{fig:ex6}, we generate a length $N =1024$ signal from a disjoint pulse stream model with $S=9$ and $F=11$; add a small amount of Gaussian noise (SNR = 13.25dB) to all its components, compute $M = 150$ noisy linear measurements, and reconstruct using Algorithm~\ref{alg:csdecon2}. The reconstructed signal is clearly a good approximation of the original signal. 

\subsection{Neuronal signals}

\begin{figure}[t]
\centering
\begin{tabular}{ccc}
{\includegraphics[width=0.30\hsize]{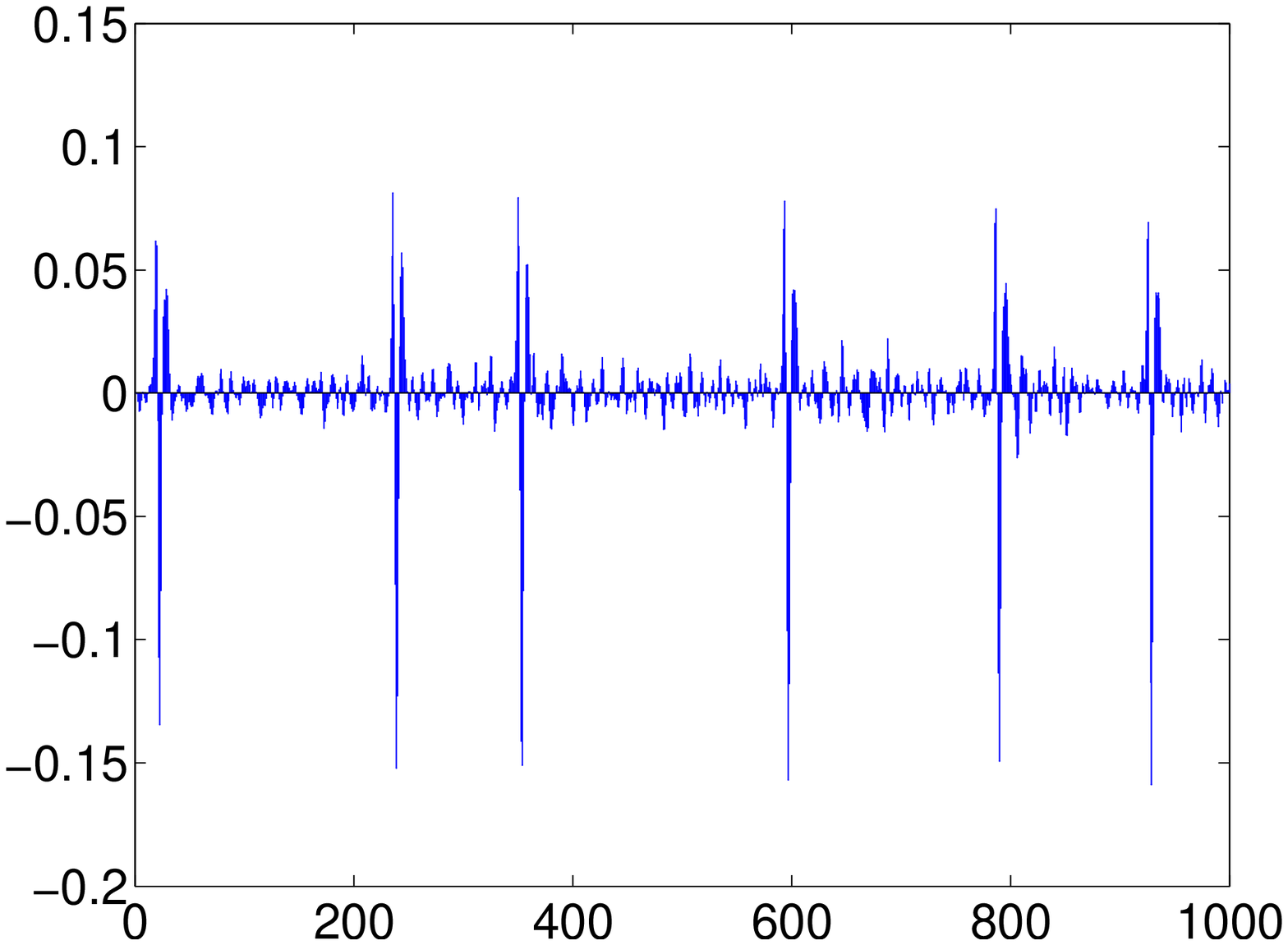}}&
{\includegraphics[width=0.30\hsize]{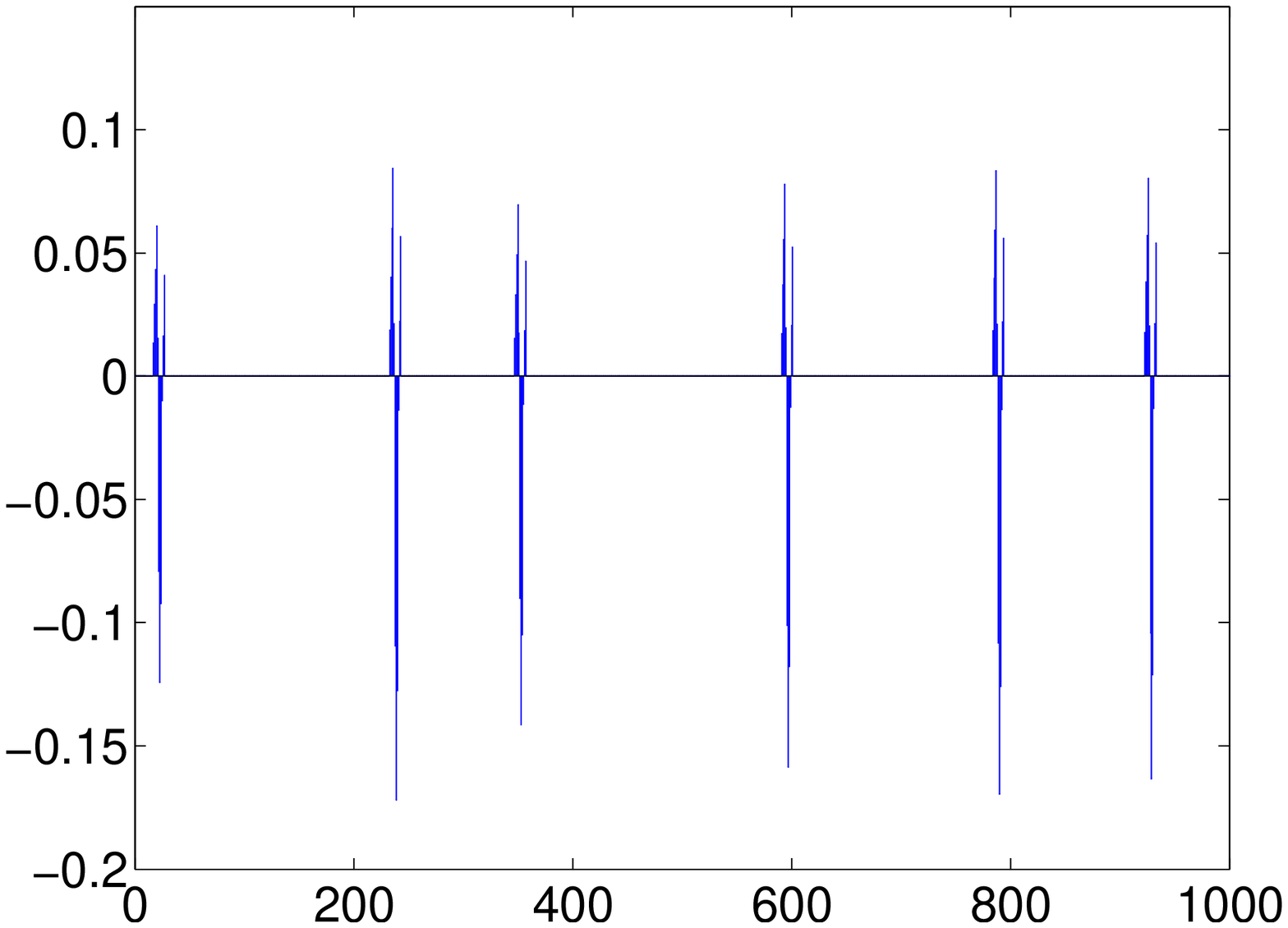}} &
{\includegraphics[width=0.30\hsize]{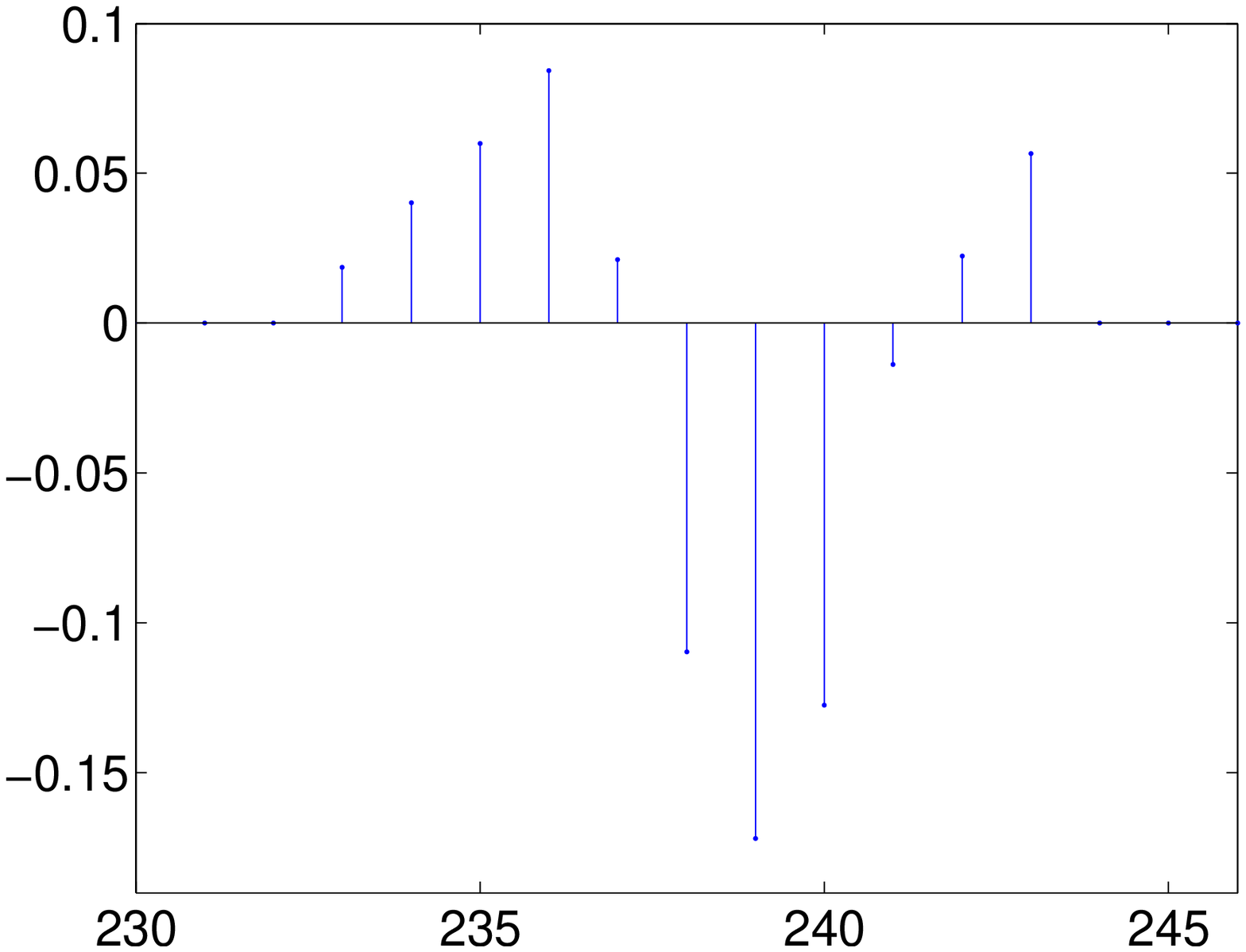}} \\
(a) & (b) & (c)
\end{tabular}
\caption{\small\sl CS recovery of a real-world neuronal signal. (a)~Original recording. (b)~Recovered signal using $M = 150$ random measurements. (c)~Estimated anchor pulse shape ($F=11$). 
\label{fig:realdata}}
\end{figure}

We test Algorithm~\ref{alg:csdecon2} on a real-world neuronal recording. Figure~\ref{fig:realdata}(a) shows the temporal electrochemical spiking potential of a single neuron. The shape of the pulses is characteristic of the neuron and should ideally be constant across different pulses. However, there exist minor fluctuations in the amplitudes, locations and profiles of the pulses. Despite the apparent model mismatch, our algorithm recovers a good approximation to the original signal (Figure~\ref{fig:realdata}(b)) as well as an estimate of the anchor pulse shape (Figure~\ref{fig:realdata}(c)). 

\subsection{Synthetic 2D pulse streams}

\begin{figure*}[t]
\centering
\begin{tabular}{ccc}
{\includegraphics[width=0.3\hsize]{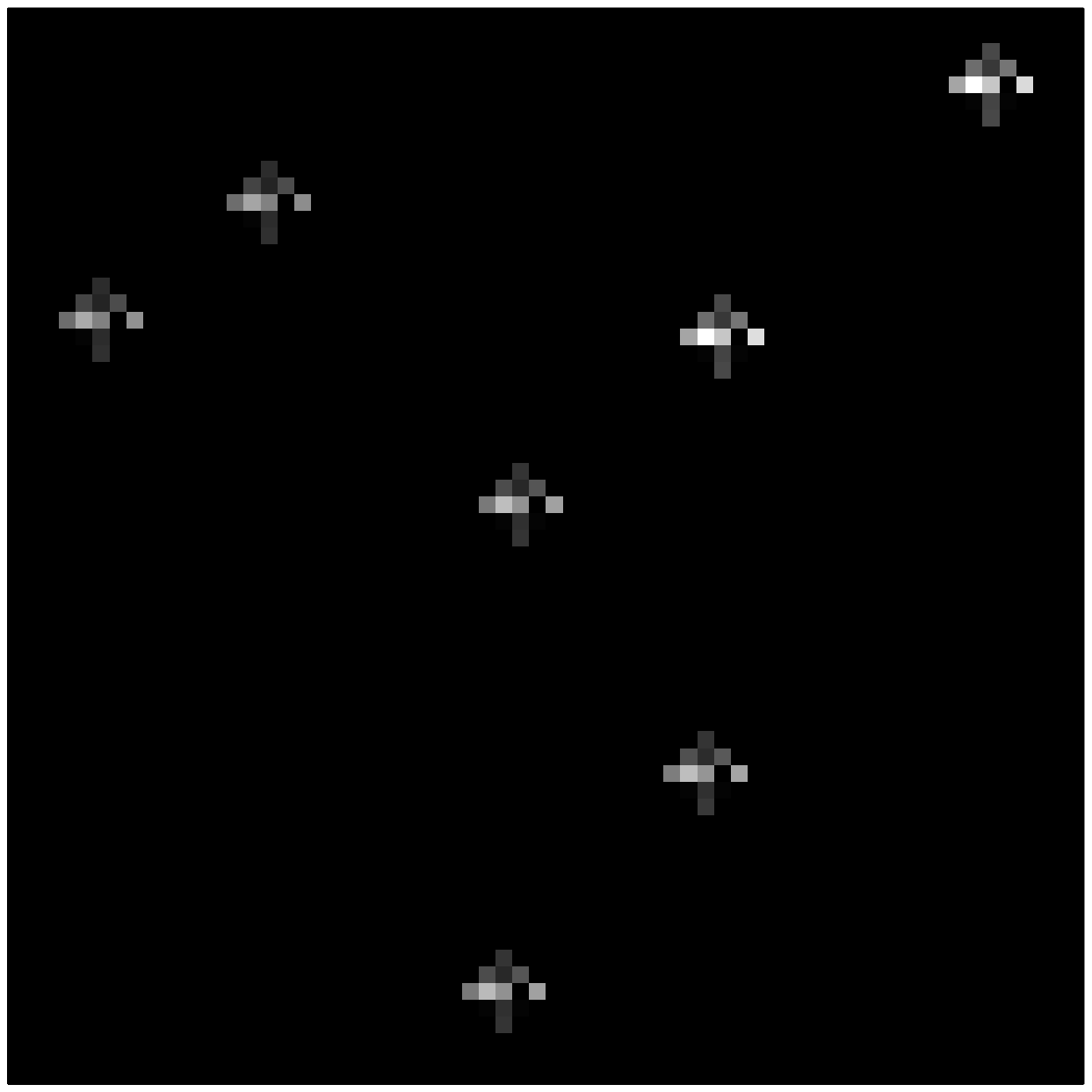}}&
{\includegraphics[width=0.3\hsize]{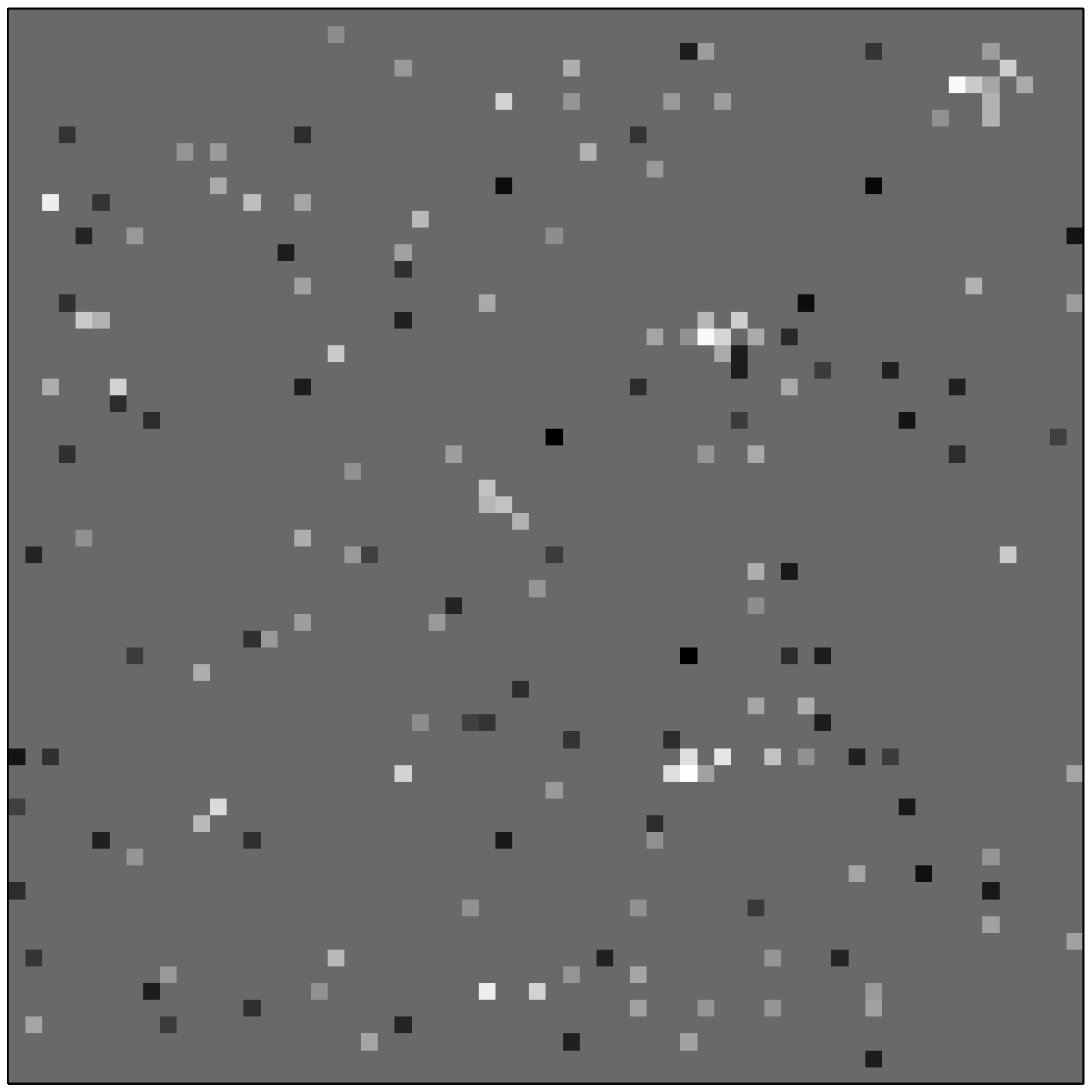}} &
{\includegraphics[width=0.3\hsize]{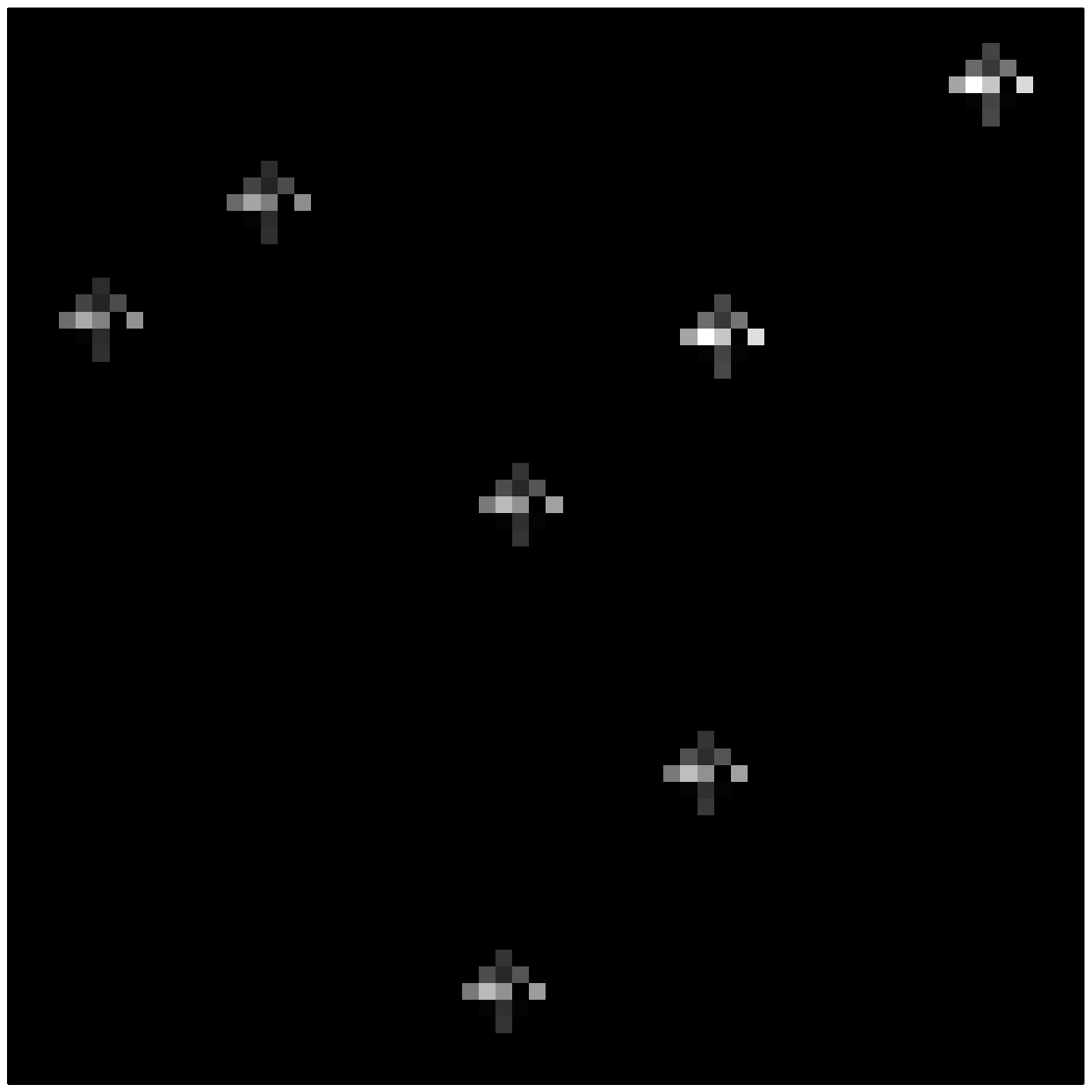}} \\
(a) & (b) & (c)
\end{tabular}
\caption{\small\sl Example CS recovery of a sum of 2D pulses. (a) Synthetic test image: $N = 4096$, $S = 7$, $F = 25$.
Images are recovered from $M=290$ random Gaussian measurements using (b) CoSaMP (MSE = 16.95), and (c) Algorithm~\ref{alg:csdecon2} (MSE = 0.07). 
\label{fig:ex2}}
\end{figure*}

Theorem~\ref{thm:zrip} and Algorithm~\ref{alg:csdecon2} can easily be extended to higher dimensional signals. For instance, suppose that the signals of interest are 2D images that can be modeled by a sparse sum of disjoint 2D pulses. We test Algorithm~\ref{alg:csdecon2} on a synthetic image (Figure~\ref{fig:ex2}(a)) of size $N = 64 \times 64 = 4096$ that comprises  $S=7$ spikes blurred by an unknown 2D impulse response of size $F = 5 \times 5 = 25$, so that the overall sparsity $K = SF = 175$. We acquire merely $M = 290$ random Gaussian measurements (approximately 7\% the size of the image $N$) and reconstruct the image using CoSaMP as well as Algorithm~\ref{alg:csdecon2}. We assume that both algorithms possess an oracular knowledge of the number of spikes $S$ as well as the size of the impulse response $F$. Figure~\ref{fig:ex2} displays the results of the reconstruction procedures using CoSaMP and Algorithm~\ref{alg:csdecon2}. It is evident both perceptually and in terms of the MSE values of the reconstructed images that our proposed approach is superior to traditional CS recovery. 

\subsection{Astronomical images}

\begin{figure*}[t]
\centering
\begin{tabular}{cc}
{\includegraphics[width=0.30\hsize]{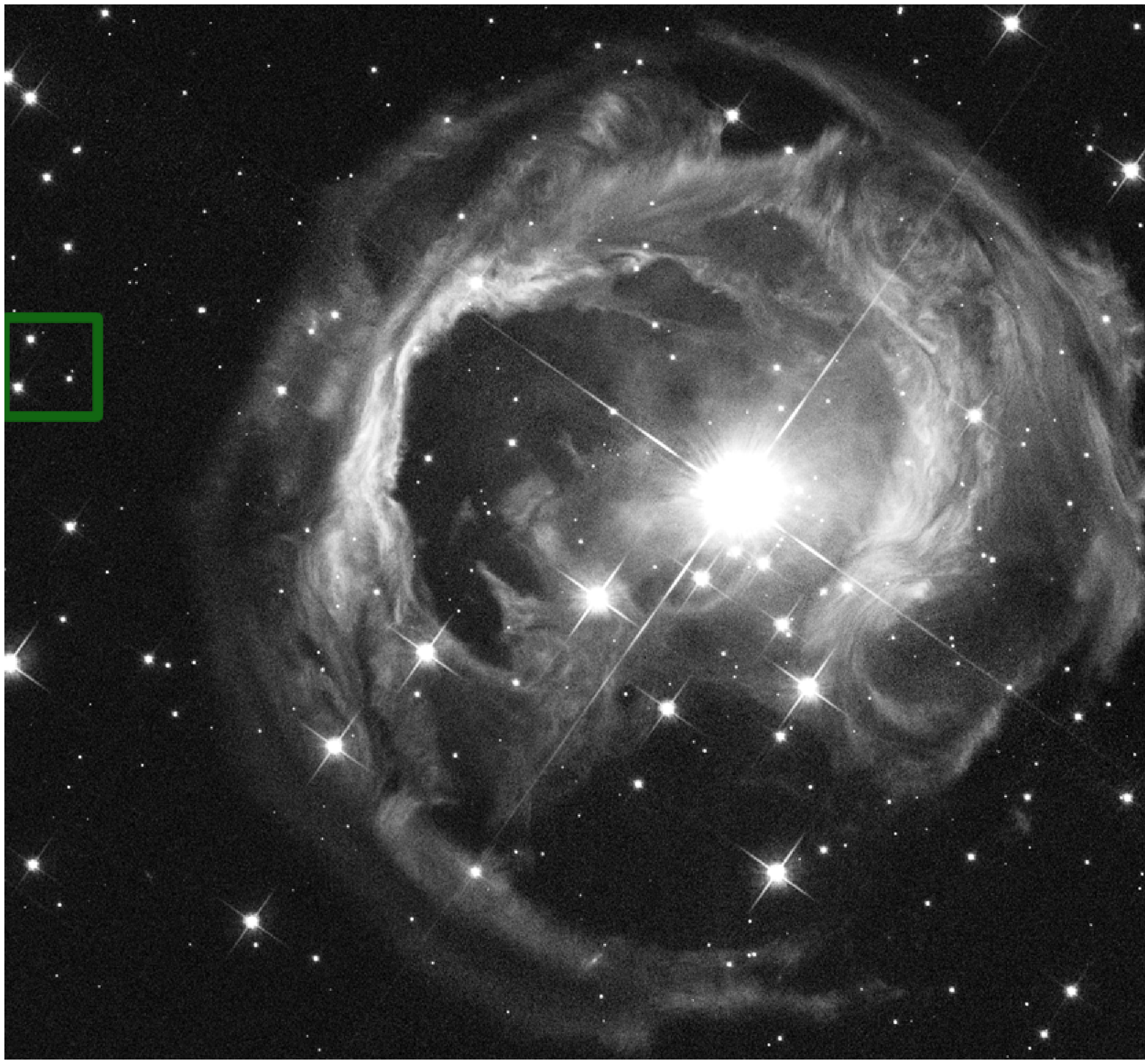}}&
{\includegraphics[width=0.30\hsize]{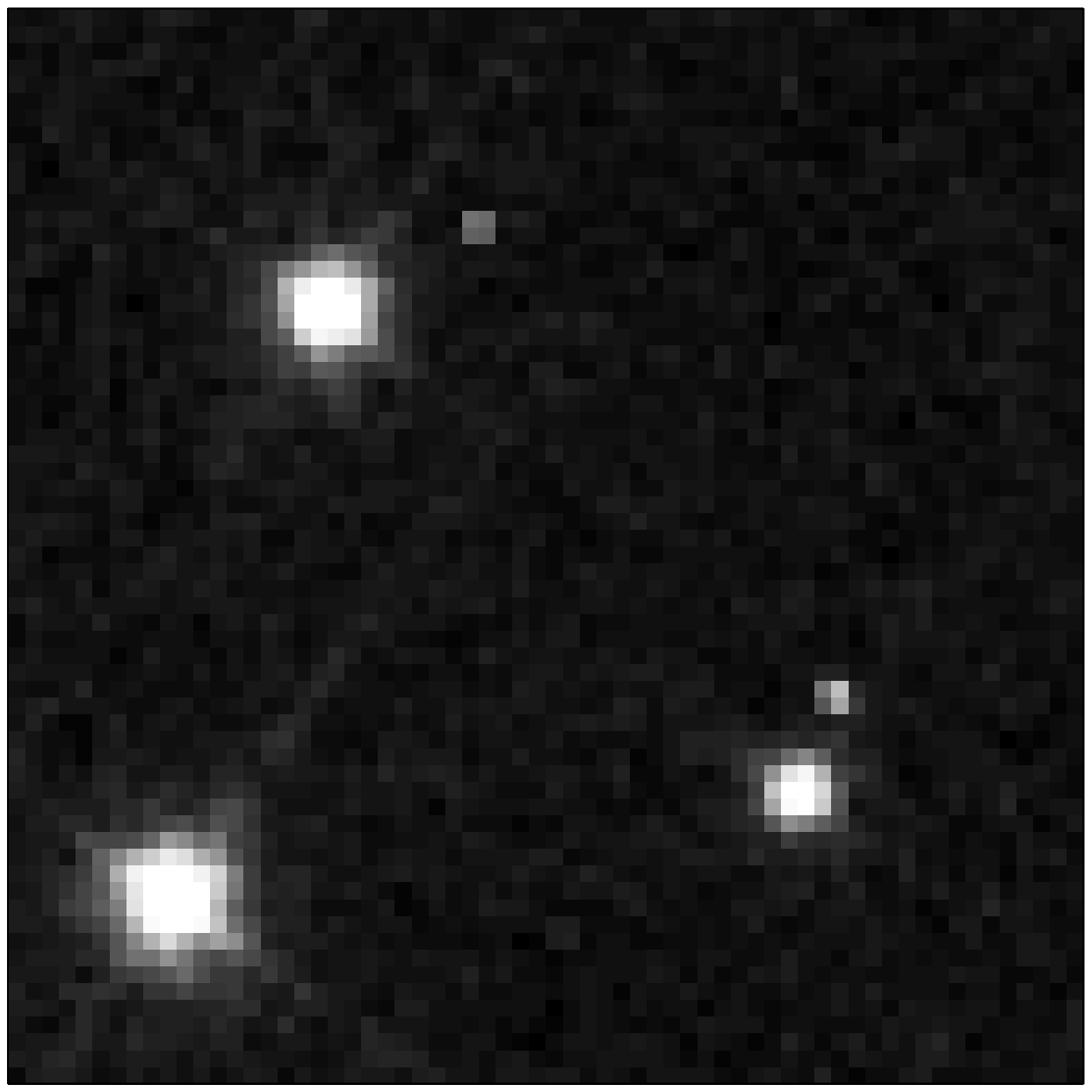}} \\
(a) & (b) \\
{\includegraphics[width=0.30\hsize]{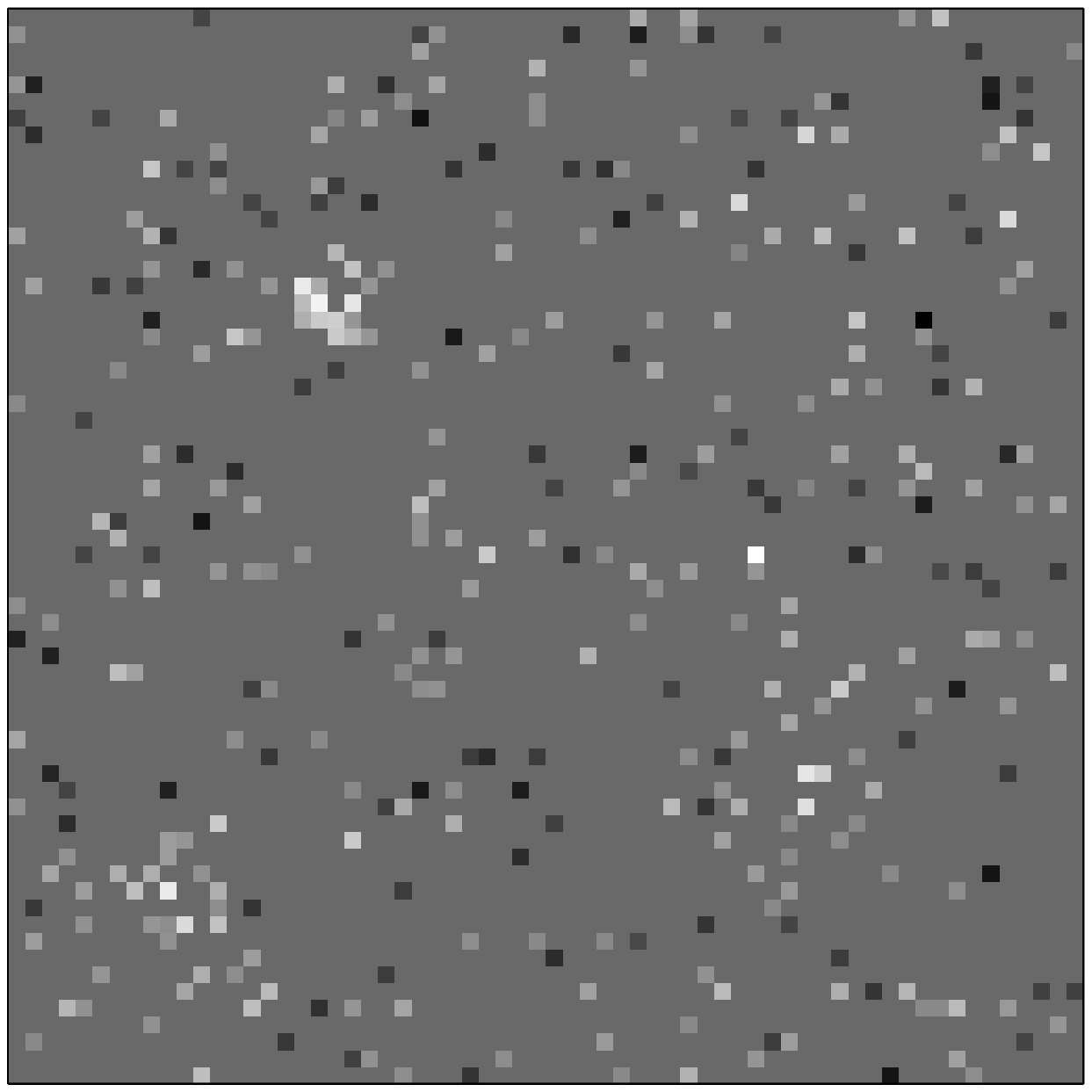}} &
{\includegraphics[width=0.30\hsize]{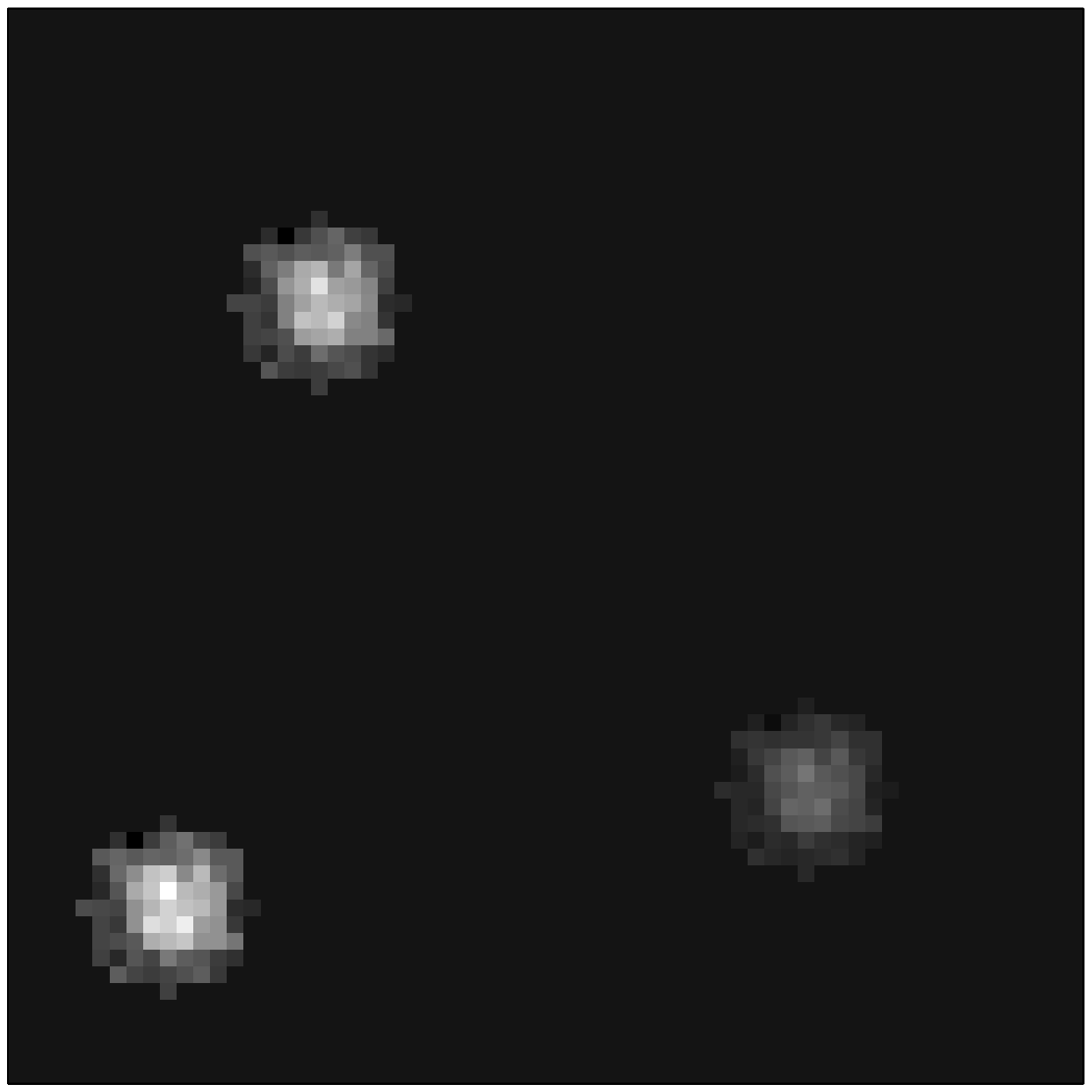}} \\
(c) & (d)
\end{tabular}
\caption{\small\sl (a) Black-and-white image of V838 Monocerotis, a nova-like star, captured by the Hubble Space Telescope on February 8, 2004~\cite{hubble}. (b) Test image is a zoomed in-version of the region highlighted in green (resolution $N = 64 \times 64 = 4096$). Reconstruction of test image is performed from $M = 330$ random Gaussian measurements using (c) CoSaMP and (d) Algorithm~\ref{alg:csdecon2}.  
\label{fig:ex7}}
\end{figure*}

Finally, we test Algorithm~\ref{alg:csdecon2} on a real astronomical image. Our test image is a $N = 64 \times 64$ region of a high-resolution image of V838 Monocerotis (a nova-like variable star) captured by the Hubble Space Telescope~\cite{hubble} (highlighted by the green square in Figure~\ref{fig:ex7}(a)). Note the significant variations in the shapes of the three large pulses in the test image (Figure~\ref{fig:ex7}(b)). We measure this image using $M = 330$ random measurements and reconstruct using both CoSaMP and Algorithm~\ref{alg:csdecon2}. For our reconstruction methods, we assumed an oracular knowledge of the signal parameters; we use $S = 3, F = 120, K = 360$ and $\Delta = 20$. As indicated by Figure~\ref{fig:ex7}, conventional CS does not provide useful results with this reduced set of measurements. In contrast, Algorithm~\ref{alg:csdecon2} gives us excellent estimates for the locations of the pulses. Further, our algorithm also provides a circular impulse response estimate that can be viewed as the anchor pulse of the three original pulses. 

\section{Discussion and Conclusions}
\label{sec:conc}
In this paper, we have introduced and analyzed a new framework for the compressive sampling of pulse streams. Our signals of interest are modeled as an infinite union of subspaces which exhibits a particular geometric structure. This structure enables us to quantitatively deduce the number of random linear measurements needed to sample such signals. We have proposed two methods for signal recovery. Our first method (Algorithm~\ref{alg:csdecon1}) is relatively easy to analyze, but suffers from combinatorial complexity. Our second method (Algorithm~\ref{alg:csdecon2}) is a feasible, if suboptimal, algorithm and formed the basis for our numerical experiments. While our framework is applicable to signals defined over domains of arbitrary dimension, we have illustrated its benefits in the context of 1D time signals and 2D images. 

There are several avenues for future work. We have discussed sparse signals and images as represented in the identity basis; our method can be extended to wavelet-sparse and Fourier-sparse signals. While our results are promising, we still do not possess a complete characterization of the convergence properties of Algorithm~\ref{alg:csdecon2} as well as its sensitivity to factors such as noise and model mismatch under random projections. Additionally, it is unclear how to deal with situations where the pulses in the signal of interest are allowed to overlap. To the best of our knowledge, the issue of robust recovery of signals convolved with an unknown arbitrary impulse response is an open question even for the case of Nyquist-rate samples. We defer these challenging open questions to future research.

The framework developed in this paper can be related to various existing concepts in the literature such as best basis compressive sensing~\cite{bestbasis}, simultaneous sparse approximation and dictionary learning~\cite{ybd}, and the classical signal processing problem of blind deconvolution~\cite{bdecon}. Compressive sensing of time-domain pulse streams has been studied by Naini {\em et al.\ }\cite{cspt}. However, in their setting the impulse response is assumed to be known, and hence the CS measurement system can be viewed as a modification of random Fourier subsampling.
 
Our framework is related to recent results on compressed blind deconvolution by Saligrama and Zhao~\cite{cbdfsp}. As opposed to pulse streams, their signals of interest consist of sparse signals driven through an all-pole auto-regressive (AR) linear system. They propose an optimization-based algorithm for recovery of the signal and impulse response from CS measurements. However, their measurement system is tailored to impulse responses corresponding to AR linear models; our approach can handle arbitrary impulse responses. Further, our main theoretical result indicates that the number of measurements to compressively sample a pulse stream is linear only in the number of degrees of freedom of the signal and thus answers an open question (Remark 3.1) posed by the authors in the affirmative.

Finally, the main approach in this paper can be related to recent work by Asif {\em et al.\ }\cite{randomcoding,randcodingbd}, who propose channel coding methods to combat the effect of unknown multipath effects in a communication channel that can be described by a sparse impulse response. Their coding strategy follows the one advocated by Cand\`{e}s and Tao~\cite{Candes04A}: their channel code consists of a random matrix $\Phi \in \real^{M \times N}$ where $M > N$, so that the linear mapping $y = \Phi x$ is now not undercomplete, but overcomplete. Thus, their observations consist of an unknown sparse channel response $h$ convolved with the transmitted signal $y$ and their objective is to reconstruct the original signal $x$. The main aspects of our theoretical analysis could conceivably be modified to quantify system performance in this setting.

\section*{Acknowledgements}
The authors would like to thank Dr.\ Volkan Cevher, Dr.\ Marco Duarte, Eva Dyer, and Mona Sheikh for helpful discussions, and Dr. Manjit Sanghera of the Dallas Presbyterian Hospital for providing the neuronal data used in Figure~\ref{fig:realdata}.

\appendices

\section{}
\label{app:jl}

We prove Theorem~\ref{thm:zrip}. By Definition~\ref{def:psm}, the model $\A_{S,F}^z$ is generated via the convolution operation by the structured sparsity models $\A_{S}$ and $\A_{F}$. Recall that both structured sparsity models are themselves defined in terms of canonical subspaces of $\real^N$ and their convolution results in a low-dimensional geometrical structure that is best described by an infinite union of subspaces.  Thus, if $x \in \A_S$ lies in a particular subspace $\Omega$ and $h \in \A_F$ lies in a particular subspace $\Lambda$, then every signal $z \in \A_{S,F}^z$ can be identified with at least one infinite union of subspaces $U_{\Omega,\Lambda}$.  The overall approach is as follows: we first construct a net of points $Q$ in $\real^N$ such that
$$
\min_{q \in Q} \| z - q \| < \delta,
$$
for all $z \in U_{\Omega,\Lambda}$ with $\| z \| = 1$ and some constant $\delta$.  We then apply the Johnson-Lindenstrauss Lemma~\cite{JL} for stable embedding of point clouds to this finite set of points $Q$, and extend the stable embedding to all possible signals $z \in U_{\Omega, \Lambda}$. Finally, we derive our main result through a union bound over all possible choices of subspaces $\Omega$ and $\Lambda$.  

Consider a fixed vector $h \in \Lambda$. Suppose the coefficients of $h$ are normalized so that $\| h \|= 1$. By virtue of its circulant nature, the spectral norm of the corresponding matrix $\| H \| \leq 1$. Now, consider a fixed $S$-dimensional subspace $\Omega \in \A_S$. It is easy to see that 
$$
\Omega_{h} = \{ z = H x ~|~ x \in \Omega \}
$$
also forms an $S$-dimensional subspace in $\real^N$. Thus, by Lemma 5.1 of~\cite{jlcs}, we can find a finite set of points $Q_{\Omega,h} \subset \Omega_h$ with cardinality $|Q_{\Omega,h}| \leq (3/\delta')^S$ such that
$$
\min_{q \in Q_{\Omega,h}} \| H x - q \| \leq \delta', ~~\forall ~\|x\| \leq 1, x \in \Omega .
$$
This is an upper bound on the size of $Q_{\Omega,h}$; assuming a worst-case scenario, we may list out the points in this set so that
$$
Q_{\Omega,h} = \{ q_1, q_2, \ldots, q_{(3/\delta')^S} \} = \{ H x_1, H x_2, \ldots, H x_{(3/\delta')^S} \} .
$$

Select any $x_l \in \{x_1, \ldots, \x_{(3/\delta')^S} \}$ and an $F$-dimensional subspace $\Lambda \in \A_F$. Form the circulant matrix $X_l$; as above, $\| X_l \| \leq 1$. Therefore, 
$$
\Omega_{x_l} = \{ z = X_l h ~|~ h \in \Lambda \}
$$
forms an $F$-dimensional subspace. Correspondingly, we can find a set of points $Q_{x_l,\Lambda} \subset \Omega_{x_l}$ with cardinality $|Q_{x_l,\Lambda}| \leq (3/\delta')^F$ such that
$$
\min_{q \in Q_{x_l,\Lambda}} \| X_l h - q \| \leq \delta', ~~\forall ~\|h\| \leq 1, h \in \Lambda .
$$
Using this process, define $Q_{x_l, \Lambda}$ for  $l = 1, 2, \ldots, (3/\delta')^S$. Then, we have
$$
Q_{\Omega, \Lambda} = \bigcup_l Q_{x_l,\Lambda} .
$$
Thus, we have identified a finite set of points $Q_{\Omega, \Lambda}$ in the infinite union of subspaces $U_{\Omega, \Lambda}$. Observe that the cardinality of this set $Q_{\Omega,\Lambda}  = (3/\delta')^S (3/\delta')^F$. Then, every vector in $U_{\Omega, \Lambda}$ with magnitude less than 1 lies `close' to at least one point in $Q_{\Omega, \Lambda}$, 
i.e., $Q_{\Omega, \Lambda}$ is a $\delta''$-net for $U_{\Omega,\Lambda}$. Suppose $\delta = 2 \delta''$. By the Johnson-Lindenstrauss Lemma, if $\Phi \in \real^{M \times N}$ with the elements of $\Phi$ drawn from a random gaussian distribution, then for every pair of vectors $z_1, z_2 \in U_{\Omega, \Lambda}$, (\ref{eq:zrip}) will hold with failure probability
$$
p_{\Omega,\Lambda} = 2 \left(\frac{3}{\delta}\right)^S \left(\frac{3}{\delta}\right)^F e^{-c_0 (\delta/2) M}.
$$
This is for a fixed pair of subspaces $(\Omega, \Lambda) \in \A_S \times \A_F$. There are $L_S \times L_F$ such pairs of subspaces. Applying a simple union bound over all possible pairs, we obtain the overall failure probability as
\begin{eqnarray*}
p&\leq& \sum_{(\Omega,\Lambda)} p_{\Omega,\Lambda} \leq L_S L_F \left(\frac{3}{\delta}\right)^{S + F} e^{-c_0 (\delta/2) M} .
\end{eqnarray*}
Rearranging terms, we have that for a suitably chosen constant $C$ (that depends on $c_0$) and for any $t > 0$, if 
$$
M \geq C \left( \log(L_S L_F) + (S + F) \log \left(\frac{3}{\delta}\right) + t \right) ,
$$
the failure probability for the sampling bound is smaller than $e^{-t}$. The theorem follows easily from this result. 

\section{}
\label{app:alg}

We prove Theorem~\ref{thm:alg1}. Let $\zhat_i = \xhat_i \ast \hhat_i$ be any intermediate estimate of Algorithm~\ref{alg:csdecon1}. Let $\widehat H_i = \mathbb{C}(\hhat_i)$. Suppose that our candidate configuration for the support of $x$ is given by the sparsity pattern $\sigma$ belonging to the structured sparsity model $\A_S^\Delta$. Then, if $(\cdot)_\sigma$ indicates the submatrix formed by the columns indexed by $\sigma$,
 the dictionary for the spike stream is given by $(\Phi \widehat H_i )_\sigma = \Phi (\widehat H_i)_\sigma$. By virtue of the least-squares property of the pseudo-inverse operator, the subsequent estimate $\xhat_{i+1}$ according to Step 2 is given by
 \begin{equation}
 \xhat_{i+1} = \arg \min_{x} \|y - \Phi (\widehat H_i)_\sigma x\|_2^2 ,
\label{eq:inf}
\end{equation}
where $x$ belongs to the $K$-dimensional subspace defined by the support configuration $\sigma$. 
Since we are minimizing a convex loss function (squared error) on a subspace in $\real^N$, the minimum $\xhat_{i+1}$ is unique. Therefore, we may view Step 2 of the algorithm as a unique-valued {\em infimal map} $f$ from a given $\hhat_i \in \A_F$ to a particular $\xhat_{i+1} \in \A_S^\Delta$. Similarly, we may view Step 4 of Algorithm~\ref{alg:csdecon1} as another unique-valued infimal map $g$ from $\A_S^\Delta$ to $\A_F$. Therefore, the overall algorithm is a repeated application of the composite map $f \circ g$. From a well-known result on single-valued infimal maps~\cite{fiorothuard,amtropp}, the algorithm is strictly monotonic with respect to the loss function. Thus, the norm of the residual $y - \Phi \zhat_i$ decreases with increasing iteration count $i$. Further, any intermediate estimate $\zhat_i$ also belongs to the model $\A(S,F,\Delta)$. 
We know from (\ref{eq:zrip}) that 
$$
\| y - \Phi \zhat_i \|_2^2 = \| \Phi z - \Phi \zhat_i \|_2^2 \geq  (1-\delta) \| z - \zhat_i \|_2^2 .
$$
Therefore, if $\| y - \Phi \zhat_i \| \leq \epsilon$, $\|z - \zhat_i \| \leq \epsilon/\sqrt{1-\delta}$.

\section{}
\label{app:anchor}

We prove Theorem~\ref{thm:anchor}. Suppose the target signal $z$ is composed of $S$ pulses $\{h_1,\ldots,h_S\}$, so that
$$
z = \widetilde{H} \widetilde{x},
$$
where $\widetilde{H} = [\mathbb{S}_1(h_1), \ldots, \mathbb{S}_S (h_S)]$ is an $N \times S$ matrix and $\widetilde{x} = (\alpha_1, \alpha_2, \ldots, \alpha_S)$. Assume we are given access to the Nyquist samples $z$, i,e., $\Phi = I_{N \times N}$. Suppose the estimate of the impulse response at an intermediate iteration is given by $\hhat$. Let $\widehat H$ be the matrix formed by the operator $\mathbb{C}(\cdot)$ acting on $\hhat$ and let $\sigma$ be the candidate support configuration for the spike stream, so that the dictionary $\Phi_h$ in this case is given by the submatrix $\hhhat_\sigma$. Note that $\hhhat_\sigma$ is quasi-Toeplitz, owing to the assumption that the separation $\Delta$ is at least as great as the impulse response length $F$. Thus, Step 2 of Algorithm~\ref{alg:csdecon1} can be represented by the least-squares operation
$$
\xhat = \widehat{H}_\sigma\pinv z .
$$
Due to the quasi-Toeplitz nature of $\widehat H_\sigma$, the pseudo-inverse $\widehat{H}_\sigma\pinv = (\hhhat_\sigma^\top \hhhat_\sigma)^{-1} \hhhat_\sigma^\top$ essentially reduces to a scaled version of the identity multiplied by the transpose of $\hhhat$ (the scaling factor is in fact the squared norm of $\hhat$). Thus, the spike coefficients are given by
$$
\xhat = \frac{1}{\| \hhat \|^2} \hhhat_\sigma^\top \widetilde{H} \widetilde{x} .
$$
Simplifying, we obtain the expression for the estimated $i^{\textrm{th}}$ spike coefficient $\alphahat_i$ as
$$
\alphahat_i = \alpha_i \frac{\langle  h_i, \hhat \rangle}{\| \hhat \|^2} .
$$
If $\hhat$ is normalized, we may write $\alphahat_i = c_i \alpha_i$, where $c_i =  \langle  h_i, \hhat \rangle$. 

Once the spike coefficients $\xhat = (c_1 \alpha_1,\ldots,c_S \alpha_S)$ have been estimated, we can form the dictionary $\Phi_x$ by considering the quasi-Toeplitz matrix $\widehat{X}$ formed by the operation $\mathbb{C}(\xhat)$. In the same manner as above, an updated estimate of the pulse shape $\widehat{\hhat}$ is given by
$$
\widehat{\hhat} = \widehat{X}\pinv z = \frac{1}{\sum_{i=1}^{S} c_i^2 \alpha_i^2 } \widehat{X}^{T} z .
$$
Writing out $\widehat{X}$ and $z$ in terms of $(h_1,\ldots,h_S)$ and $(\alpha_1,\ldots,\alpha_S)$ and simplifying, we obtain
$$
\widehat{\hhat} = \frac{ \sum_{i=1}^S c_i \alpha_i^2 h_i} {\sum_{i=1}^S c_i^2 \alpha_i^2} ,
$$
where $c_j = \langle  h_j, \hhat \rangle$. Thus, we have a closed-expression for the updated estimate of the impulse response coefficients $\widehat{\hhat}$ in terms of the previous estimate $\hhat$. In the event that the algorithm converges to a fixed point, we can replace $\widehat{\hhat}$ by $\hhat$, thus proving the theorem.

{{
\bibliography{CSBib,chin,ramp-ciss}
\bibliographystyle{ieeetr}
}}

\end{document}